\documentclass[12pt,nofootinbib,superscriptaddress]{revtex4}

\usepackage{amssymb,color,paralist,amsmath,epsfig}
\usepackage{graphicx}
\usepackage{amsfonts}
\usepackage{nicefrac,esint}
\usepackage{verbatim}
\usepackage{float}

\usepackage{relsize}

\usepackage[colorlinks,citecolor=blue,linktoc=all,linkcolor=cyan]{hyperref}
\usepackage[section]{placeins}

\newcommand{\cG}{{\cal G}}
\newcommand{\cF}{{\cal F}}

\newcommand{\hp}{{\frac{1}{2}}}
\newcommand{\hm}{{-\frac{1}{2}}}

\newcommand{\beq}{\begin{equation}}
\newcommand{\eeq}{\end{equation}}

\newcommand{\ber}{\begin{eqnarray}} 
\newcommand{\eer}{\end{eqnarray}}

\usepackage{color}

\usepackage[normalem]{ulem}

\renewcommand\sout{\bgroup \color[rgb]{0,0.00,1.} \ULdepth=-.5ex \ULset}

\begin{document}

\title{Two-photon exchange corrections to elastic $e^-$-proton scattering: Full dispersive treatment of $\pi N$ states at low momentum transfers}
\author{Oleksandr Tomalak}
\affiliation{Institut f\"ur Kernphysik and PRISMA Cluster of Excellence, Johannes Gutenberg Universit\"at, Mainz, Germany}
\author{Barbara Pasquini}
\affiliation{Dipartimento di Fisica, Universit\`a degli Studi di Pavia, Pavia, Italy}
\affiliation{INFN Sezione di Pavia, Pavia, Italy}
\author{Marc Vanderhaeghen}
\affiliation{Institut f\"ur Kernphysik and PRISMA Cluster of Excellence, Johannes Gutenberg Universit\"at, Mainz, Germany}

\date{\today}

\begin{abstract}
We evaluate the pion-nucleon intermediate-state contribution to the two-photon exchange (TPE) correction in the elastic electron-nucleon scattering within a dispersive framework. We calculate the contribution from all $\pi N$ partial waves using the MAID parametrization. We provide the corresponding TPE correction to the unpolarized $e p$ scattering cross section in the region of low momentum transfer $ Q^2 \lesssim 0.064~\mathrm{GeV}^2$, where no analytical continuation into the unphysical region of the TPE scattering amplitudes is required. We compare our result in the forward angular region with an alternative TPE calculation, in terms of structure functions, and find a good agreement, indicating a small contribution at low $Q^2$ due to discontinuities beyond $ \pi N$. We also compare our results with empirical fits.

\end{abstract}

\maketitle

\vspace{-0.2cm}

\tableofcontents

\section{Introduction}
\label{sec1}

Since the first electron scattering experiments off nucleons and nuclei at the Stanford University High-Energy Physics Laboratory more than six decades ago by Hofstadter and his team ~\cite{hofs53, hofs55}, 
lepton scattering has become the prime source of information on the internal structure of strongly interacting systems and has provided such fundamental nucleon and nuclear structure quantities as 
their density distributions through elastic and inelastic form factors, their polarizabilities, their detailed parton distributions, and various generalizations thereof. 
The success of the electroweak probe as a source of structure information 
is mainly due to the perturbative nature of the interaction, which allows one to 
describe such scattering processes to good approximation as due to the exchange of one photon, or one weak vector boson between the lepton and the nucleon or nucleus. This allows one to cleanly separate the interaction of the probe from the response functions of the target, which encode the relevant information of the nucleon or nucleus structure.
With the increasing precision of such scattering experiments and with the technical realization of experiments with polarized leptons and/or polarized nucleons,  one is presently able to test the limitations of the simple one-photon exchange (OPE) description of the probing interactions. A first clear signal came 
at the dawn of the present  century 
from a series of measurements  at the Jefferson Lab (JLab) of the electric over magnetic proton form-factor ratio $G_{Ep} /  G_{Mp}$ through either the scattering of polarized electrons off polarized protons or the measurement of the recoiling proton's polarization. These experiments have yielded the unexpected result that with increasing values of the momentum transfer $Q^2$, the ratio $G_{Ep} /  G_{Mp}$ shows an approximate linear decrease~\cite{Jones:1999rz, Gayou:2001qd, Punjabi:2005wq, Puckett:2010ac}, in clear contradiction with the near-constant ratio obtained from unpolarized experiments; see Ref.~\cite{Punjabi:2015bba} for a recent review. 
The first explanations of such behavior pointed towards hard two-photon exchange (TPE) processes between the electron and the proton, which become relevant once experiments aim to access terms which contribute at or below the percent level to the scattering cross section~\cite{Guichon:2003qm, Blunden:2003sp}. 
Such observations have triggered new experiments specifically designed to measure the TPE effects, 
and have opened a whole field of theoretical calculations aimed to estimate TPE effects; see Refs.~\cite{Carlson:2007sp, Arrington:2011dn} for some reviews. 

On the experimental side, there exist observables which provide us with very clear indications of the size of TPE effects, as they would be exactly zero in the absence of two- or multiphoton-exchange contributions. Such observables are normal single-spin asymmetries (SSA) of electron-nucleon scattering, where either the electron spin or the nucleon spin is polarized normal to the scattering plane. Because such SSAs are proportional to the imaginary part of a product of two amplitudes, they are zero for real (nonabsorptive) processes such as OPE. At leading order in the 
fine-structure constant, $\alpha \simeq 1/137$, they result from the product of the OPE amplitude and the imaginary part of the TPE amplitude. 
For the target normal SSA, they were predicted to be in the (sub) percent range some time ago~\cite{DeRujula:1972te}. Recently, a first measurement  of the normal SSA for the elastic 
electron-$^3$He scattering  has been performed by the JLab Hall A Collaboration, 
extracting a SSA for the elastic electron-neutron subprocess in the percent range~\cite{Zhang:2015kna}.  
For the experiments with polarized beams, the corresponding normal SSAs were predicted to be in the range of a few to hundred ppm for electron beam energies in the GeV range~\cite{Afanasev:2002gr, Gorchtein:2004ac, Pasquini:2004pv}. Although such asymmetries are small, the parity-violation programs at the major electron laboratories have reached precisions on asymmetries with longitudinal polarized electron beams well  below the ppm level, and the next generations of such experiments are designed to reach precisions at the sub-ppb level~\cite{Kumar:2013yoa}. The beam normal SSA, which is due to TPE and thus parity conserving,  
has been measured over the past fifteen years as a spinoff by the parity-violation experimental collaborations at MIT-BATES (SAMPLE Collaboration)~\cite{Wells:2000rx}, 
at MAMI (A4 Collaboration)~\cite{Maas:2004pd, BalaguerRios:2012uk}, and at JLab (G0 Collaboration~\cite{Armstrong:2007vm, Androic:2011rh}, HAPPEX/PREX Collaboration~\cite{Abrahamyan:2012cg}, and Qweak Collaboration~\cite{Waidyawansa:2013yva, Nuruzzaman:2015vba}).  The resulting beam normal SSA ranges from a few ppm in the forward angular range to around a hundred ppm in the backward angular range, in qualitative agreement with theoretical TPE expectations. While the nonzero normal SSAs in elastic electron-nucleon scattering quantify the 
imaginary parts of the TPE amplitudes, measurements of their real parts have also been performed by several dedicated experiments over the past few years. In particular, the deviation from unity of the  elastic scattering cross-section ratio $R_{2 \gamma} \equiv e^+ p / e^- p$ is proportional to the real part of the product of OPE and TPE amplitudes. Recent measurements of $R_{2 \gamma}$, for $Q^2$ up to $2$~GeV$^2$,  
have been performed at VEPP-3~\cite{Rachek:2014fam},  
by the CLAS Collaboration at JLab~\cite{Adikaram:2014ykv, Rimal:2016toz}, and by the  
OLYMPUS Collaboration at  DESY~\cite{Henderson:2016dea}. These experiments show that $R_{2 \gamma}$ 
ranges,  for the kinematical region corresponding with $Q^2 = 0.5 - 1$~GeV$^2$  
and virtual photon polarization parameter $\varepsilon = 0.8 - 0.9$,  
from a value $R_{2 \gamma} \approx 0.99$~\cite{Henderson:2016dea}, showing a deviation from unity within $2 - 3~ \sigma$ (statistical and uncorrelated systematic errors),  
to a value $R_{2 \gamma} = 1.02 - 1.03$ for $Q^2 \approx 1.5$~GeV$^2$  and $\varepsilon \approx 0.45$~\cite{Rachek:2014fam, Rimal:2016toz}. 
Furthermore, the  GEp2gamma Collaboration at JLab has performed a pioneering measurement of the deviation from the OPE prediction in two double-polarization observables of the ${\vec e} p \to e \vec p$ process at 
$Q^2 = 2.5$~GeV$^2$~\cite{Meziane:2010xc}, and has found a deviation from the OPE result in one of these observables (corresponding with longitudinal recoil proton polarization) at the 4$\sigma $ level at $\varepsilon = 0.8$.  
In combination with $R_{2 \gamma}$, measurements of the $\varepsilon$ dependence of these two double-polarization observables in the ${\vec e} p \to e \vec p$ process at a fixed value of $Q^2$ allow us to experimentally fully disentangle the TPE amplitudes for massless electrons~\cite{Guttmann:2010au}.  
 
A good quantitative understanding of TPE corrections to the lepton-proton amplitude also became of paramount  
importance in recent years in the interpretation of Lamb shift and hyperfine splitting measurements of muonic Hydrogen ($\mu$H) and muonic atoms. For the $2P_{1/2} - 2S_{1/2}$ $\mu$H Lamb shift (LS), a very accurate measurement has been obtained~\cite{Pohl:2010zza, Antognini:1900ns}: 
\begin{equation}
\Delta E_{\mathrm{LS}}^{\mathrm{exp}} (\mu \mathrm{H}) = 202.3706 (23)~\mathrm{meV}. 
\end{equation}
The theoretical calculation of this transition for the $\mu$H system displays a very strong sensitivity to the radius of the proton charge distribution $r_E$ as~\cite{Antognini:2013jkc} 
\begin{equation}
\Delta E_{\mathrm{LS}}^{\mathrm{th}} (\mu \mathrm{H}) = 206.0336 (15) - 5.2275 (10) r_E^2 + \Delta E_{\mathrm{TPE}},
\label{eq:LS}
\end{equation}
with $E$ in meV and $r_E$ in fm.   
The first (dominant) term is due to the well-known (mainly) QED contributions, while the second contains the dependence on $r_E^2$ which one likes to extract. While the muonic system has a very strong sensitivity  to the proton charge distribution due to a  third-power dependence in the lepton mass, it simultaneously also 
displays a strong sensitivity to the next-order (in $\alpha$) hadronic corrections, denoted by 
 $\Delta E_{\mathrm{TPE}}$ in Eq.~(\ref{eq:LS}). 
This term corresponds with the TPE process between the muon and the proton at $Q^2 \approx 0$, 
which is also known as a polarizability correction.   
Dispersive estimates, connecting this quantity to forward proton structure functions, have yielded $\Delta E_{\mathrm{TPE}} (\mu \mathrm{H})  = 0.0332 (20)$~meV \cite{Carlson:2011zd, Birse:2012eb, Antognini:2013jkc}. One thus notices that although the size of the TPE correction is only around $1.6 \times 10^{-4}$ of the total 
$2P - 2S$ LS for $\mu$H, its uncertainty is of the same size as the experimental uncertainty on 
$\Delta E_{\mathrm{LS}}$ 
and is at present the main theoretical uncertainty when converting $\Delta E_{\mathrm{LS}}$ 
to a value for the proton radius $r_E$. 
This situation is even enhanced when considering muonic deuterium ($\mu$D). In this case the 
TPE correction amounts to around $8.4 \times 10^{-3}$ of the total $2P - 2S$ Lamb shift~\cite{Pohl1:2016xoo, Krauth:2015nja}. In the nuclear case the experimental uncertainty 
$\delta \Delta E_{\mathrm{LS}}^{\mathrm{exp}} (\mu \mathrm{D}) = 3.4~\mu$eV~\cite{Pohl1:2016xoo}  is dwarfed by the present 
theoretical uncertainty of the TPE correction 
$\delta \Delta E_{\mathrm{TPE}} (\mu \mathrm{D}) = 20.0~\mu$eV~\cite{Krauth:2015nja}. As the TPE corrections  are at present the main uncertainty in the extraction of the charge radii of the proton and light nuclei from LS measurements, 
this calls for new efforts at model-independent approaches in calculating TPE corrections. Such efforts are underway by evaluating them within chiral effective field theory see Refs.~\cite{Pineda:2004mx,Nevado:2007dd} for the first computation with chiral effective theories for the Lamb shift and Ref.~\cite{Hagelstein:2015egb} for a review of the ongoing activity in this field; within non-relativistic QED~\cite{Pineda:2002as,Hill:2012rh, Dye:2016uep}; or by connecting them model-independently to other data through dispersive frameworks. 

The muonic atom spectroscopy has made it very apparent that there is an urgent need for a reevaluation of TPE corrections, going beyond model-dependent assumptions made in earlier works. But even within the accuracy of the present calculations, these measurements have also revealed for the LS splittings in the $\mu$H and $\mu$D cases that the extracted values for the proton and deuteron charge radii are at variance with the values extracted from electron scattering or electronic hydrogen 
spectroscopy~\cite{Bernauer:2010wm, Bernauer:2013tpr, Mohr:2012tt}. Proton   
charge radius extractions, differing by as much as 7$\sigma$, have led to the proton radius puzzle~\cite{Bernauer:2014cwa}, which has resulted lately in a lot of activity to understand its origin; 
see Ref.~\cite{Carlson:2015jba} for a recent review. 
One of the avenues of new work is to extend the electron scattering experiments to lower values of $Q^2$, 
in order to reduce  uncertainties in the radius extraction resulting from an extrapolation in $Q^2$ to the region below the  lowest values which were accessed in   
Refs.~\cite{Bernauer:2010wm, Bernauer:2013tpr}. Such new experiments are currently being performed at MAMI 
(ISR experiment ~\cite{Mihovilovic:2016rkr}) as well as at JLab (PRad experiment~\cite{Peng:2016szv}), which both aim at reaching $Q^2$ values down to 10$^{-4}$~GeV$^2$. Furthermore, elastic scattering experiments using not only electron beams, but also muon beams are planned at PSI (MUSE experiment~ \cite{Gilman:2013eiv}) to cross-check the lepton universality in the extracted proton form factors, and to test the size of the TPE effects by comparing $e^+/e^-$ and $\mu^+/\mu^-$ results. All these new experimental programs  aim at subpercent-level precision on the elastic cross sections, necessitating a corresponding improved quantitative understanding of the hadronic corrections, in particular the radiative and TPE corrections.
 
Over the past decade, a large number of model estimates have been made for hard TPE contributions to elastic lepton-proton scattering,  after an initial hadronic box diagram estimate with nucleon intermediate states only~\cite{Blunden:2003sp}, and a quark-based estimate in terms of integrals over 
generalized parton distributions~\cite{Chen:2004tw, Afanasev:2005mp}. 
In the very large $Q^2$ region (relative to the hadronic mass scale), one may resort to perturbative techniques to estimate TPE 
corrections within QCD factorization frameworks~\cite{Borisyuk:2008db, Kivel:2009eg, Kivel:2012vs}. 
In the very small $Q^2$ region, the leading TPE correction is given by the term of order $Q$ and can be obtained as a Coulomb limit~\cite{McKinley:1948zz} of the elastic contribution. The leading inelastic TPE results in a $Q^2 \ln Q^2$ correction term to the cross section, which may be expressed  as a weighted integral over the photoabsorption cross section on a 
nucleon~\cite{Brown:1970te, Gorchtein:2014hla}. The subleading correction terms in $Q^2$ can be obtained 
through corresponding integrals over the $Q^2$-dependent proton structure functions, 
parametrizing the forward doubly virtual Compton scattering 
process~\cite{Gorchtein:2006mq, Tomalak:2015aoa, Tomalak:2015hva}. 
When studying TPE corrections in the low-to-intermediate range of $Q^2$ (of the order of a hadronic scale around 1 GeV$^2$) however, one is in the nonperturbative regime of QCD, where the relevant degrees of freedom are hadronic.  
In this regime, box-diagram models with nucleons have only been extended to include contributions from inelastic intermediate states. 
In particular, the inclusion of the $\Delta$(1232) resonance, being the lowest nucleonic excitation in the electroproduction process, has been studied by 
several groups~\cite{Kondratyuk:2005kk, Borisyuk:2012he, Graczyk:2013pca, Zhou:2014xka, Lorenz:2014yda}. Even though some of these calculations give results with a pathological behavior in the $\varepsilon \to 1$ limit (at fixed $Q^2$), signaling a violation of unitarity, these works qualitatively point towards an increasing  
$\Delta$ TPE contribution with larger $Q^2$ and with decreasing values of  $\varepsilon$ (at a fixed $Q^2$).   
Within hadronic models, higher intermediate states have also been included through box-diagram-type calculations, pointing towards a partial cancellation between the contributions from spin-1/2 and spin-3/2 resonances~\cite{Kondratyuk:2007hc}. 

Although several hadronic models may provide reliable estimates for the imaginary parts of TPE amplitudes, where only the on-shell (i.e., experimentally directly accessible) information is used, the calculation of the real parts requires off-shell information, which introduces model dependence.  At present, there are two types of theoretical approaches  which conceptually can provide this information in a way consistent with unitarity and analyticity. The first are effective field theories (EFTs) based on a power-counting scheme. A recent example is a chiral EFT calculation with $\pi$, $N$, and $\Delta$ degrees of freedom of the TPE correction to the $\mu$H Lamb shift~\cite{Peset:2014jxa,Alarcon:2013cba}. The second are dispersive formulations which allow one to go beyond $\pi$, $N$, and $\Delta$ degrees of freedom and may also be extended to larger values of $Q^2$. 
Its input are on-shell electroproduction amplitudes for $N$, $\pi N$, $\pi \pi N$, ... intermediate states, which fix the imaginary parts of the TPE and can be tested through the measurements of normal SSAs. The real parts of the TPE amplitudes are then reconstructed through a dispersion relation. Such calculations have been performed for the nucleon intermediate state, exploiting unsubtracted and subtracted dispersion relations in Refs. \cite{Borisyuk:2008es,Tomalak:2014sva}. 
The extension of the dispersive formalism to calculate the TPE amplitude with a spin-3/2 $\Delta$ intermediate state and with spin-1/2 and spin-3/2 intermediate states has been considered in Refs.~\cite{Borisyuk:2012he, Borisyuk:2013hja, Borisyuk:2015xma}. In the latter works, the empirical multipoles for pion electroproduction were reparametrized as a sum of monopole form-factor behaviors in the virtualities of each of the two photons, effectively simplifying the calculation to one-loop box diagrams.  

\begin{figure}[h]
\begin{center}
\includegraphics[width=.44\textwidth]{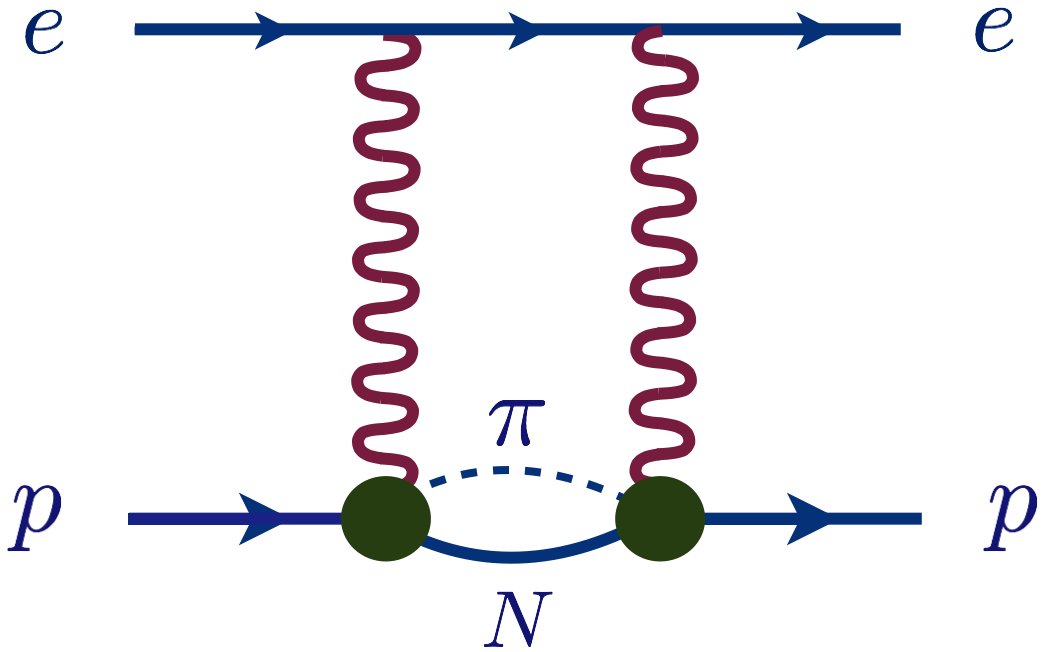}
\end{center}
\caption{TPE graph with $ \pi N $ intermediate state.}
\label{piN_TPE}
\end{figure}

It is the aim of the present work to present a dispersive treatment of all $\pi N$ intermediate-state contributions to the TPE amplitudes in elastic electron-proton scattering; see Fig. \ref{piN_TPE}, which directly uses as input    
the most recent pion electroproduction amplitudes from the MAID2007 fit~\cite{Drechsel:1998hk,Drechsel:2007if}. 
For the imaginary parts of the TPE amplitudes, our work builds upon a previous study performed in Ref.~
\cite{Pasquini:2004pv}.  
Our work goes beyond previous dispersive estimates, as it accounts for all known intermediate states with spins 1/2, 3/2, 5/2, 7/2,... A complete study is particularly relevant in view of cancellations between spin-1/2 and spin-3/2 states found earlier in Ref. \cite{Borisyuk:2015xma}. Furthermore, a correct account of nonresonant t-channel pion and vector-meson exchange contributions in the pion electroproduction process requires us to go beyond the lowest few partial waves. In addition, we formulate the problem such that the invariant amplitudes for pion electroproduction can be directly used as input, without having to approximate the analytic structure of the $\gamma^\ast N \to R$ vertices by sums of monopoles. 
In particular, for the $\gamma^\ast N \to \Delta$ vertex, chiral EFT calculations~\cite{Pascalutsa:2005ts, Pascalutsa:2005nd, Pascalutsa:2006up} have indicated  that the corresponding form factors 
display important branch-cut contributions due to pion loops, providing $Q^2$ dependencies beyond a simple monopole behavior. 
In the present work, we provide the dispersive evaluation of the cross-section correction at low momentum transfer $ Q^2 < 0.064~\mathrm{GeV}^2$, which corresponds with the value below for which no analytical continuation of imaginary parts into the unphysical region for the TPE amplitudes is required. We also provide a detailed comparison of all partial wave contributions in MAID with the {\it near-forward} approximation of Ref.~\cite{Tomalak:2015aoa}. At low momentum transfer, such a comparison quantifies the contributions beyond the leading inelastic $\pi N$ channel. Using both unsubtracted and subtracted DR formalisms and accounting for the $ \pi N $ intermediate states, we provide first quantitative estimates of the full contribution of the $ \pi N $ channel to the inelastic TPE correction, and compare this with empirical TPE fits at low $Q^2$ values. 
In a subsequent paper, we plan to present the extension of our work to larger values of $Q^2$. For $Q^2 > 0.064~\mathrm{GeV}^2$, we will outline a procedure how to analytically continue the TPE amplitudes based on phenomenological $ \pi N$ input.

The plan of the present paper is as follows: We describe the general formalism of the TPE corrections in the elastic electron-proton scattering in Sec. \ref{sec2}. The $ \pi N $ contribution to the imaginary parts of TPE amplitudes and to the target normal single spin asymmetry are evaluated from unitarity relations in Sec. \ref{sec3}. The $ \pi N $ contribution to the real parts of TPE amplitudes are evaluated using dispersive integrals in Sec. \ref{sec4}. We present the cross-section correction, compare the unsubtracted DR evaluation with the {\it near-forward} approximation, and discuss the subtracted DR formalism in Sec. \ref{sec5}. Our conclusions and outlook are given in Sec. \ref{sec6}. The pion electroproduction process is described in two appendixes.

\section{TPE correction in elastic $ ep$ scattering}
\label{sec2}

\begin{figure}[h]
\begin{center}
\includegraphics[width=.45\textwidth]{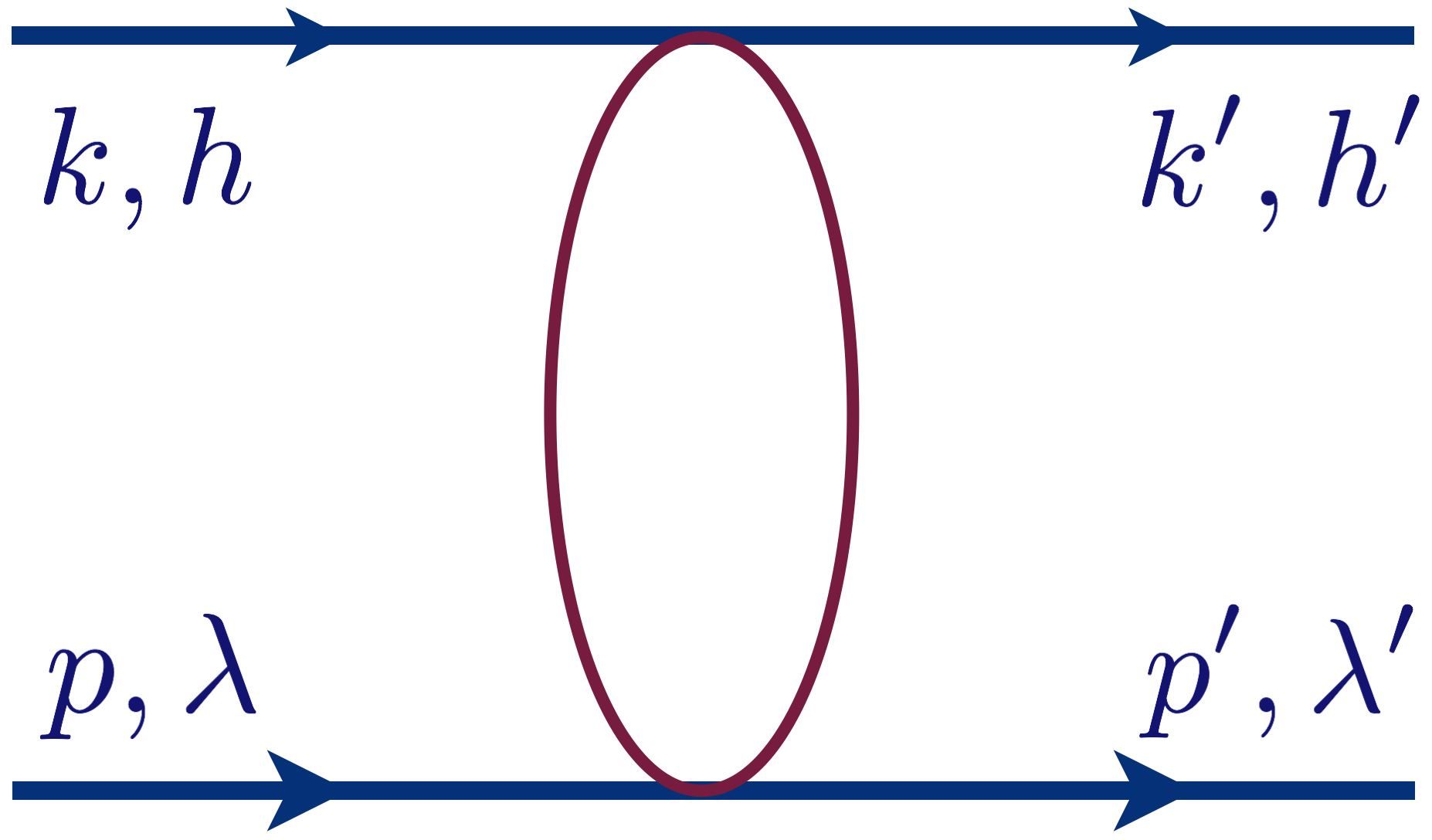}
\end{center}
\caption{Elastic electron-proton scattering.}
\label{elastic_scattering_general}
\end{figure}

We consider in this work the elastic electron-proton scattering process: $ e( k , h ) + p( p, \lambda ) \to e( k^\prime, h^\prime) + p(p^\prime, \lambda^\prime) $,  where the variables in brackets denote the four-momenta and helicities, rerspectively,  of the participating particles; see Fig. \ref{elastic_scattering_general}. This process is completely described by two Mandelstam variables, e.g., $ Q^2 = - (k-k^\prime)^2 $ the squared momentum transfer, and $ s = ( p + k )^2 $ the squared total energy in the electron-proton center-of-mass (c.m.) reference frame.

In terms of the laboratory-frame four-momenta $ p = (M, 0), ~ k = (\omega, \vec{k}), ~ k^\prime = (\omega^\prime, \vec{k}^\prime), ~ p^\prime = (E_p^\prime, \vec{k}-\vec{k}^\prime)$, the invariant variables are expressed as 
\ber
Q^2 = 2 M ( \omega - \omega^\prime) , \quad s = M^2 + 2 M \omega,
\eer
with the proton mass $M$. To exploit the symmetry with respect to $ s\leftrightarrow u $ crossing, where $ u = ( k - p^\prime )^2 $, it is convenient to introduce a crossing-symmetric kinematical variable as $ \nu \equiv ( s - u )/4= M \omega - Q^2 /4$. The virtual photon polarization parameter $ \varepsilon $, which is often used in analyzing $e^{-}p$ scattering data, can be expressed in terms of $ Q^2 $ and $ \nu $ as
\ber
\label{epsilon}
\varepsilon & = & \frac{  \nu^2 - M^4 \tau_P ( 1 + \tau_P )}{  \nu^2 + M^4 \tau_P ( 1 + \tau_P )} ,
\eer
where $ \tau_P = Q^2/ (4 M^2)$. It varies between 0 and 1, indicating the degree of longitudinal polarization of the virtual photon in the case of one-photon exchange.

The momentum transfers explored so far in elastic $e^- p$ scattering experiments, 
$ Q^2 \gtrsim 10^{-4} ~ \mathrm{GeV}^2 $, are much larger than the squared electron mass. 
Consequently, electrons can be treated as massless particles. Assuming the conservation of parity 
and time-reversal invariance, the helicity amplitudes for the elastic $ e^{-} p $ scattering (for 
massless electrons) can be parametrized in terms of three independent Lorentz-invariant 
amplitudes that are complex functions of two independent kinematical variables~\cite{Guichon:2003qm}:
\begin{eqnarray}
 \label{str_ampl} 
T_{h^\prime \lambda^\prime, h \lambda}  &=&  
\frac{e^2}{Q^2} \bar{u}(k^\prime,h^\prime) \gamma_\mu u(k,h) \nonumber\\
&\times & \bar{N}(p^\prime,\lambda^\prime) 
 \left( \gamma^\mu \cG_M (\nu, Q^2) -   \frac{P^{\mu}}{M} \cF_2 (\nu, Q^2) 
 +  \frac{\gamma . K P^{\mu}}{M^2} 
  \cF_3 (\nu, Q^2) \right)N(p,\lambda),
  \label{str_ampl1}
\end{eqnarray}
where the averaged momentum variables are $ P = (p+p^\prime)/2, ~ K = (k+k^\prime)/2 $; $u$ ($ \bar{u}$) is the initial (final) electron spinor; $N$ ($ \bar{N}$) is the initial (final) proton spinor; and $\gamma. a  \equiv \gamma^\mu a_\mu$. In the approximation of one exchanged photon, $ \cF^{1 \gamma}_3 = 0 $, and the remaining two independent amplitudes are equivalent to the magnetic $ G_M (Q^2) = \cG_M^{1 \gamma} $ and electric $ G_E (Q^2) = G_M ( Q^2 )- ( 1 + \tau_P) F_2 (Q^2) $ proton form factors, where $ F_2 (Q^2) = \cF_2^{1 \gamma} $ is the Pauli form factor.

Only three helicity amplitudes are independent, e.g. $ T_1 \equiv T_{\hp \hp, \hp \hp}$, $ T_2 \equiv T_{\hp \hm, \hp \hp}$ and $T_3 \equiv T_{\hp \hm, \hp \hm} $ in the c.m. reference frame. Following the Jacob-Wick \cite{Jacob:1959at} phase convention, the invariant amplitudes can be expressed through the helicity amplitudes as \cite{Pasquini:2004pv}
\ber \label{ff_s}
e^2 \cG_M & = & \frac{1}{2} \left\{ T_1 - T_3 \right\}, \nonumber \\
e^2 \cF_2 & = & \frac{ M Q }{\sqrt{ M^4 - s u}} \left\{ - T_2 + T_3 \frac{ M Q }{\sqrt{ M^4 - s u}} \right\}, \nonumber \\
e^2 \cF_3 & = & \frac{M^2}{s - M^2} \left\{ - T_1 - T_2 \frac{ 2 M Q }{\sqrt{ M^4 - s u }} + T_3 \left( 1 + Q^2\frac{ s + M^2 }{ M^4 - s u }\right) \right\}. 
\eer

In the following, we will alternatively work also with the amplitudes $ \cG_1,~\cG_2 $, defined as
\ber 
 {\cal{G}}_{1}  & = & {\cal{G}}_{M}  + \frac{\nu}{M^2} {\cal{F}}_3 = \frac{Q^2}{e^2 \left(s - M^2 \right)} \left( \frac{T_1 + T_3}{4} - \frac{M}{Q} \frac{2 \nu }{\sqrt{ M^4 - s u}} T_2 \right) + \frac{ M^2 Q^2 \left( 1 + \tau_P \right) }{ M^4 - s u} \frac{T_3}{e^2},  \label{new_amplitude1}  \\ 
 {\cal{G}}_{2}  & = & {\cal{G}}_{M}  - \left( 1 + \tau_P \right) {\cal{F}}_2  + \frac{\nu}{M^2} {\cal{F}}_3 = \frac{Q^2}{4 e^2 \left(s - M^2 \right)} \left( T_1 + T_3 + \frac{Q}{M} \frac{ s + M^2 }{\sqrt{ M^4 - s u}} T_2 \right) . \label{new_amplitude2}
\eer

The TPE correction $ \delta_{2 \gamma} $, at leading order in $ \alpha \equiv e^2 / 4 \pi \simeq 1/137$,  is defined through the difference between the cross section accounting for the exchange of two photons $ (\sigma) $ and the cross section in the $1 \gamma $-exchange approximation $ (\sigma_{1 \gamma}) $ as
\ber \label{TPE_definition}
 \sigma \equiv \sigma_{1 \gamma} \left( 1 + \delta_{2 \gamma} \right).
\eer
In terms of the invariant amplitudes, the TPE correction to the unpolarized $ e^- p $ cross section at the leading $ \alpha $ order is given by
\beq \label{unpolarized_cross_section}
\delta_{2 \gamma} = \frac{2}{G^2_M  + \frac{\varepsilon}{\tau_P} G^2_E } \left\{ G_M  \Re  {\cal{G}}_{1}^{2 \gamma}  + \frac{\varepsilon}{\tau_P} {G_E}  { \Re  \cal{G}}_{2}^{2 \gamma}  +  {G_M} \left (\varepsilon -1\right) \frac{\nu}{M^2} {{\Re \cal{F}}_3^{2 \gamma}}  \right\}.
\eeq

In the following, we will also discuss the target normal spin asymmetry which vanishes in the one-photon-exchange approximation and therefore allows us to cross-check the theoretical TPE calculations. The target normal spin asymmetry ($A_n$) resulting from the scattering of unpolarized electrons on protons polarized normal to the scattering plane (with the proton spin $ S = \pm S_n $) is defined as \cite{DeRujula:1972te,Pasquini:2004pv}
\ber \label{TSSA_def}
A_n = \frac{\mathrm{d} \sigma \left( S = S_n \right) - \mathrm{d} \sigma \left( S = -S_n \right)}{\mathrm{d} \sigma \left(  S = S_n \right) + \mathrm{d} \sigma \left( S = -S_n \right)}.
\eer
The asymmetry in Eq. (\ref{TSSA_def}) is expressed through the imaginary parts of the TPE amplitudes at the leading $ \alpha $ order as
\ber \label{TSSA}
A_n & = & \sqrt{\frac{2 \varepsilon \left( 1 + \varepsilon \right)}{\tau_P}} \frac{1}{G^2_M + \frac{\varepsilon}{\tau_P} G^2_E } \left\{ - G_M \Im \cG^{2 \gamma}_2 + G_E \Im \left( \cG^{2 \gamma}_1 - \frac{1-\varepsilon}{1+\varepsilon} \frac{\nu}{M^2}  \cF^{2 \gamma}_3 \right) \right \}.
\eer

In Eqs. (\ref{unpolarized_cross_section}, \ref{TSSA}), the index $ 2 \gamma $ indicates the TPE contribution to the invariant amplitudes.

\section{Unitarity relations}
\label{sec3}

\begin{figure}[h]
\begin{center}
\includegraphics[width=1.\textwidth]{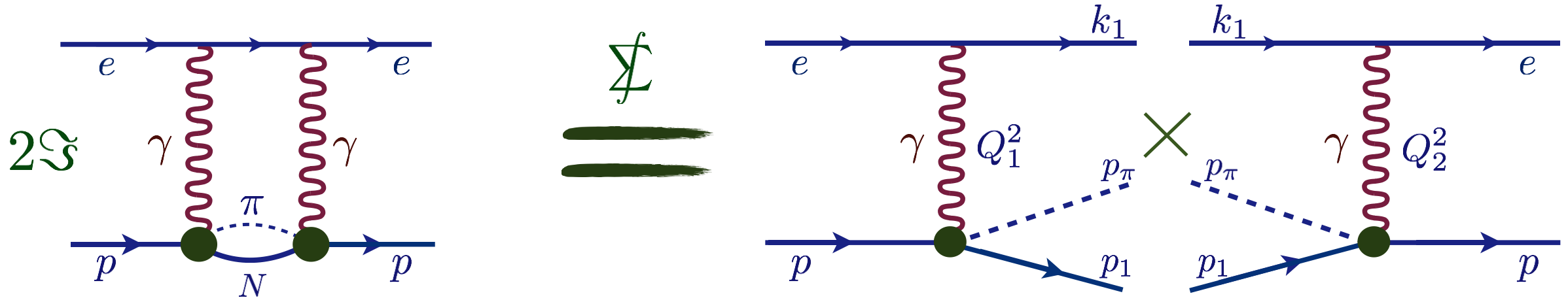}
\end{center}
\caption{Unitarity relations for the case of the $ \pi N $ intermediate-state contribution.}
\label{unitarity_piN}
\end{figure}

In this work, we evaluate the TPE contribution to the elastic electron-proton scattering arising from the inelastic $ \pi N $ channel (see Fig. \ref{piN_TPE}) using a dispersive formalism. The first step in this formalism is the calculation of the imaginary parts of the invariant amplitudes. They can be obtained with the help of the unitarity equation for the $S$-matrix:
\ber
 S^+ S  =  1,\qquad T^+ T = i ( T^+ - T),
\eer
where the $ T $-matrix is defined as $ S = 1 + i ~T $.
In the case of the $ \pi N $ intermediate state with the pion ($ p_\pi $), proton ($ p_1 $), and electron ($ k_1 $) intermediate momenta,
\ber
p_\pi = (E_\pi, \vec{p}_\pi ), \qquad p_1 = (E_1, \vec{p}_1 ), \qquad k_1 = (\omega_1, \vec{k}_1 ),
\eer
the unitarity relation gives for the imaginary part of the TPE helicity amplitudes~\cite{Pasquini:2004pv}
\ber \label{piN_unitarity}
\Im T^{2 \gamma}_{h^\prime \lambda^\prime, h \lambda}  = \frac{e^4}{2} \int \frac{\mathrm{d}^3 \vec{k}_1}{ (2 \pi)^3 2 \omega_1}\frac{1}{Q^2_1 Q^2_2} \bar{u}(k^\prime,h^\prime) \gamma_\mu \gamma.k_1 \gamma_\nu u(k,h) \cdot \bar{N}(p^\prime,\lambda^\prime) W^{\mu \nu} \left( p, p^\prime, k_1   \right) N(p,\lambda), \nonumber \\
\eer
where $ Q^2_1=-(k-k_1)^2$ and $Q^2_2=-(k^\prime-k_1)^2$ are the virtualities of the two spacelike  photons, as indicated in Fig. \ref{unitarity_piN}. Furthermore, the hadronic tensor $ W^{\mu \nu}$ is defined as
\ber \label{piN_hadronic_tensor}
W^{\mu \nu} \left( p, p^\prime, k_1    \right)  =  \int \frac{\mathrm{d}^3 \vec{p}_1 }{(2 \pi)^3 2 E_1} \int \frac{d^3 \vec{p}_\pi }{(2 \pi)^3 2 E_\pi} && \hspace{-0.4cm} (2 \pi)^4 \delta^4(k + p - k_1 - p_1 - p_\pi) \nonumber \\
&& \hspace{-0.32cm} \times \tilde{J}^{\mu}_{\pi N}(p_\pi, p_1, p^\prime )   \left( \gamma.p_1 + M \right)  J_{\pi N}^\nu (p_\pi, p_1, p),
\eer
where  $J_{\pi N}^{\nu}$ and $\tilde{J}^{\mu}_{\pi N}$ are  the pion electroproduction currents describing the excitation and deexcitation of the $\pi N$ intermediate states, respectively~\cite{Dennery:1961zz,Berends:1967vi,Pasquini:2007fw}. As we aim in this work at an empirical estimate of the inelastic TPE contribution due to the $ \pi N $ channel, we take the MAID 2007 parametrization \cite{Drechsel:1998hk,Drechsel:2007if} as input in calculating the pion production currents; see Appendix \ref{app2} for details. \footnote{Note that the elastic contribution is obtained by replacing the pion production current in the hadronic tensor of Eq. (\ref{piN_hadronic_tensor}) with the proton current and changing the phase space integration as follows:
\ber
&& J_{\pi N}^\mu (p_\pi, p_1, p)  \to  G_M  \gamma^\mu - F_2  \frac{p^{\mu} + p^\mu_1}{2 M}, \nonumber
\\
&&\int \frac{\mathrm{d}^3 \vec{p}_1 }{(2 \pi)^3 2 E_1} \int \frac{d^3 \vec{p}_\pi }{(2 \pi)^3 2 E_\pi} (2 \pi)^4 \delta^4(k + p - k_1 - p_1 - p_\pi)  \to  2 \pi \delta(W^2 - M^2).\nonumber
\eer
}

\begin{figure}[h]
\begin{center}
\includegraphics[width=0.77\textwidth]{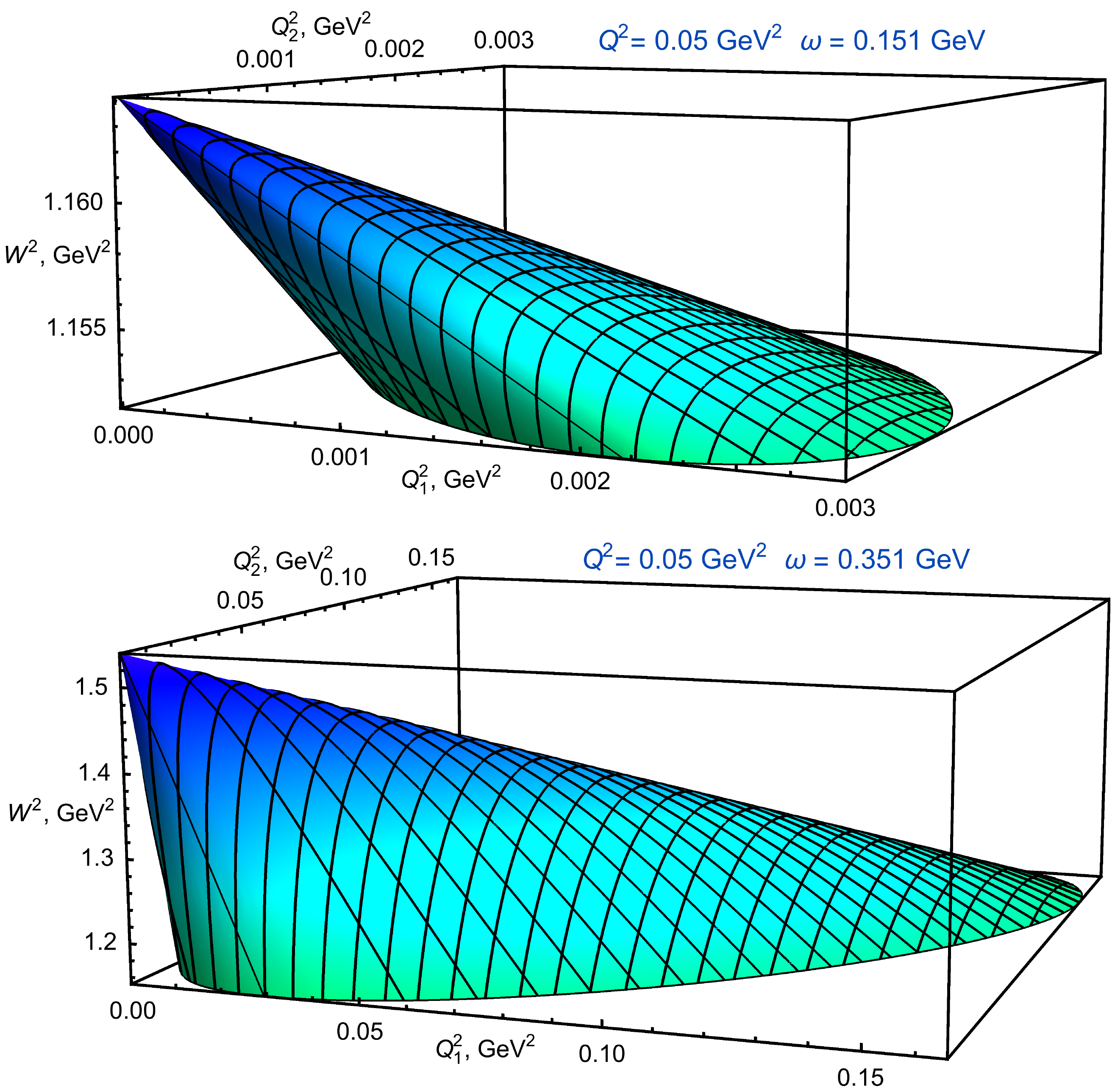}
\end{center}
\caption{Integration region in Eq. (\ref{unitarity_pion_state}) as a function of the  virtualities of the exchanged photons and the invariant mass of the intermediate $\pi N$ system.
The results are for $Q^2=0.05$ GeV$^2$ and two different values of the electron beam energy:
 $ \omega = 0.151~\mathrm{GeV}$, corresponding with $\varepsilon = 0.2 $ (upper panel), and $ \omega = 0.351~\mathrm{GeV}$, corresponding with $\varepsilon = 0.8$ (lower panel).}
\label{inelastic_ellipse}
\end{figure}

The integral in Eq. (\ref{piN_unitarity}) runs over the three-momentum of the intermediate (on-shell) electron. In the $e^- p$ c.m. system, the momentum of the intermediate electron reads
\ber
|\vec k_1|=\frac{s-W^2}{2\sqrt{s}},
\eer
where $ W^2 = \left( p_1 + p_\pi \right)^2 $ is the squared invariant mass of the intermediate pion-nucleon system. 
Using $W^2$ as the variable of integration, 
the integral (\ref{piN_unitarity}) can be rewritten as
\ber 
&&\Im T^{2 \gamma}_{h^\prime \lambda^\prime, h \lambda}  =   e^4 \int  \limits^{s}_{(M+m_\pi)^2}   \mathrm{d} W^2  \int \mathrm{d} \Omega_1 \frac{   \omega_1 }{64 \pi^3 {\sqrt{s}} Q^2_1 Q^2_2}  \nonumber \\
&& \hspace{4cm} \times \bar{u}(k^\prime,h^\prime) \gamma_\mu \gamma.k_1 \gamma_\nu u(k,h) \cdot \bar{N}(p^\prime,\lambda^\prime) W^{\mu \nu} \left( p, p^\prime, k_1 \right) N(p,\lambda).\label{unitarity_pion_state}
\eer 
In Eq. \eqref{unitarity_pion_state}, we define the polar  c.m.  angle $\theta_1$ of the intermediate electron with respect to the direction of the initial electron, and the azimuthal angle $\phi_1$ is chosen in such a way that $\phi_1=0$ corresponds with the scattering plane of the $ep\rightarrow ep$ process.
The virtualities of the exchanged photons can be expressed as
\beq \label{Q21_Q22_ellipse}
 Q^2_1 =  \frac{ \left( s - W^2 \right) \left( s - M^2 \right)}{2s} \left( 1 - \cos \theta_1 \right), ~~~ Q^2_2 =   \frac{ \left( s - W^2 \right) \left( s - M^2 \right)}{2s}  \left( 1 - \cos \theta_2 \right),
\eeq
where $\theta_2$ is the  angle between the intermediate and final electrons.
In terms of the polar and azimuthal angles $\theta_1$ and $\phi_1$ of the intermediate electron, and the c.m. scattering angle of the final electron $\theta_{c.m.}$, one has
\ber
\cos\theta_2=\sin\theta_{c.m.}\sin\theta_1\cos\phi_1+\cos\theta_{c.m.}\cos\theta_1.
\eer
Therefore, the two-dimensional integral over the solid angle $\Omega_1$ is  equivalent to an integral over the ellipse in the $ Q^2_1,~Q^2_2$ plane defined by Eq. (\ref{Q21_Q22_ellipse}).
The  corresponding full three-dimensional integration regions in Eq. (\ref{unitarity_pion_state})  are shown in Fig.  \ref{inelastic_ellipse} for two kinematics used in this work.

The hadronic tensor in Eq. (\ref{unitarity_pion_state}) is given by
\ber
W^{\mu \nu} \left( p, p^\prime, k_1    \right) & = &  \int \frac{  \mathrm{d} \Omega_\pi }{(4 \pi)^2 }  \frac{ |\vec{p}_\pi|^2 \tilde{J}_{\pi N}^{\mu} (p_\pi, p_1, p^\prime)   \left( \gamma.p_1 + M \right)  J_{\pi N}^\nu \left(p_\pi, p_1, p\right)}{ |\vec{p}_\pi| \left( \sqrt{s} - \omega_1 \right) + E_\pi \omega_1 ( \hat{k}_1 \cdot \hat{p}_\pi)},
\eer
where $ \hat{k}_1 \equiv \vec{k}_1/|\vec{k}_1|$ and $ \hat{p}_\pi \equiv \vec{p}_\pi/|\vec{p}_\pi|$. We note that in the kinematical region defined by
\ber
 \left[ W^2-(M +m_\pi)^2 \right] \left[W^2-(M -m_\pi)^2 \right] \le 4 m^2_\pi \omega_1^2,
\eer
with the pion mass $ m_\pi$, we have to sum over the two possible solutions for $ |\vec{p}_\pi|$. See Appendix \ref{app1} for details of the pion electroproduction kinematics. 

After performing the numerical integration for the helicity amplitudes in Eq. (\ref{unitarity_pion_state}), the imaginary parts of the TPE invariant amplitudes at  leading order in $\alpha$ are then obtained by using Eq. (\ref{ff_s}). 

The pion electroproduction amplitudes from MAID \cite{Drechsel:1998hk,Drechsel:2007if}, which we use to evaluate Eq. (\ref{unitarity_pion_state}),  are available in the restricted kinematical region $ W < W_{\mathrm{MAID}} = 2.5 ~\mathrm{GeV} $. The {\it near-forward} approximation of Ref. \cite{Tomalak:2015aoa} is based on the unpolarized structure functions, which are available in MAID in the region $ W < W_{\mathrm{MAID}} = 2 ~\mathrm{GeV} $. When performing the dispersion integrals in Sec. \ref{sec4}, we will integrate over the whole phase space in the unitarity relation of Eq. (\ref{unitarity_pion_state}), when the crossing symmetric variable satisfies:
\ber
  \nu < \nu_{\mathrm{MAID}} = \frac{W^2_{\mathrm{MAID}} - M^2}{2} - \frac{Q^2}{4}.
 \eer 
Note that $ \nu_{\mathrm{MAID}} \approx 2.67 ~\mathrm{GeV}^2  $ ($ \nu_{\mathrm{MAID}} \approx 1.55 ~\mathrm{GeV}^2  $) at $ Q^2 = 0.05 ~\mathrm{GeV}^2 $ and $ \nu_{\mathrm{MAID}} \approx 2.68 ~\mathrm{GeV}^2  $ ($ \nu_{\mathrm{MAID}} \approx 1.56 ~\mathrm{GeV}^2  $) at $ Q^2 = 0.005 ~\mathrm{GeV}^2 $, for $ W_{\mathrm{MAID}} = 2.5 ~\mathrm{GeV} $ and $W_{\mathrm{MAID}} = 2 ~\mathrm{GeV}$, respectively. For larger $ \nu > \nu_{\mathrm{MAID}} $, we will truncate the $ W $ integration in Eq. (\ref{unitarity_pion_state}) at $ W = W_{\mathrm{MAID}} $ (instead of $ W = \sqrt{s}$), accounting only for the available kinematical region in MAID. Consequently, the imaginary parts of the invariant amplitudes will show a kink at the point $ \nu = \nu_{\mathrm{MAID}} $. In Sec. \ref{sec5}, we will estimate the residual contribution from the region $ W > W_{\mathrm{MAID}}$.

\begin{figure}[htp]
\begin{center}
\includegraphics[width=0.81\textwidth]{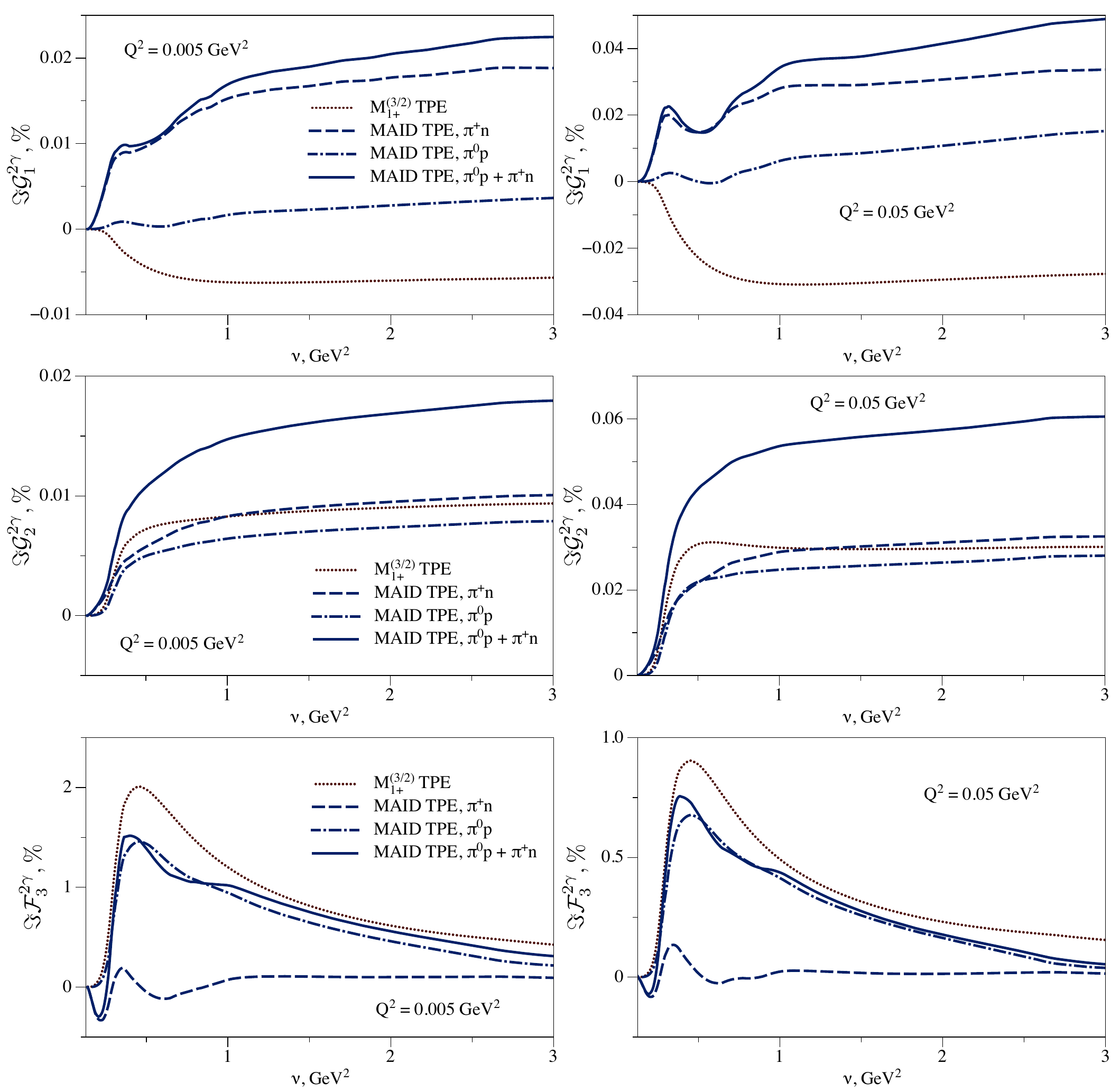}
\end{center}
\caption{Imaginary part of the inelastic $\pi N$ contribution to the TPE amplitudes $ \cG^{2 \gamma}_1,~\cG^{2 \gamma}_2,~\cF^{2 \gamma}_3$  in the elastic electron-proton scattering for $ Q^2 = 0.005 ~\mathrm{GeV}^2 $ (left panels) and $ Q^2 = 0.05 ~\mathrm{GeV}^2 $ (right panels), calculated from the MAID solutions of the pion electroproduction amplitudes. 
The different curves show the  contributions from the $M^{(3/2)}_{1+}$ multipole, and  
from all the  partial waves in the $ \pi^0 p$ and $ \pi^+ n$ channels separately, and that from the sum of the $ \pi^0 p$ and $ \pi^+ n $ channels (see the legend for the notations).} 
\label{P33_imaginary_G1_G2_F3}
\end{figure}

As the $ \Delta$(1232) resonance is the most prominent $ \pi N $ resonance, corresponding to the $P_{33} $ $\pi N$ partial wave, \footnote{$ L_{2\mathrm{I} 2\mathrm{J}} $ is the partial wave with the pion angular momentum
$ L $, isospin $ \mathrm{I} $, and total angular momentum $ \mathrm{J} $.} we will also separately evaluate the contribution from the $ \Delta$ resonance, restricting ourselves to the dominant magnetic dipole $ M^{(3/2)}_{1+}$ with the isospin $ \mathrm{I} = 3/2 $ in the following. \footnote{$ M^{(\mathrm{I})}_{\mathrm{L} \mathrm{\pm}} $ is the multipole with the pion angular momentum
 $ \mathrm{L} $, isospin $ \mathrm{I} $, and total angular momentum $ \mathrm{J} = \mathrm{L} \pm \frac{1}{2} $.}
In Fig.  \ref{P33_imaginary_G1_G2_F3}, we present the results for the imaginary parts of the 
electron-proton TPE amplitudes $ \cG^{2 \gamma}_{1},~\cG^{2 \gamma}_{2},~\cF^{2 \gamma}_{3}$ 
for $ Q^2 = 0.005 ~\mathrm{GeV}^2 $ and $ Q^2 = 0.05 ~\mathrm{GeV}^2$, using MAID as input.
We show the contributions from the $M^{(3/2)}_{1+}$ multipole, and from all $\pi N$ partial waves in the 
$ \pi^0 p $  and $ \pi^+ n $ channels separately, as well as that from the sum of the $ \pi^0 p$ and $ \pi^+ n $ channels. \footnote{ Note that the projection of the $M^{(3/2)}_{1+}$ multipole into the $ \pi^0 p$ and $ \pi^+ n$ channels is equal to $2/3$ and $1/3$ of the total result, respectively, as dictated by isospin symmetry.}

\begin{figure}[t]
\begin{center}
\includegraphics[width=0.62\textwidth]{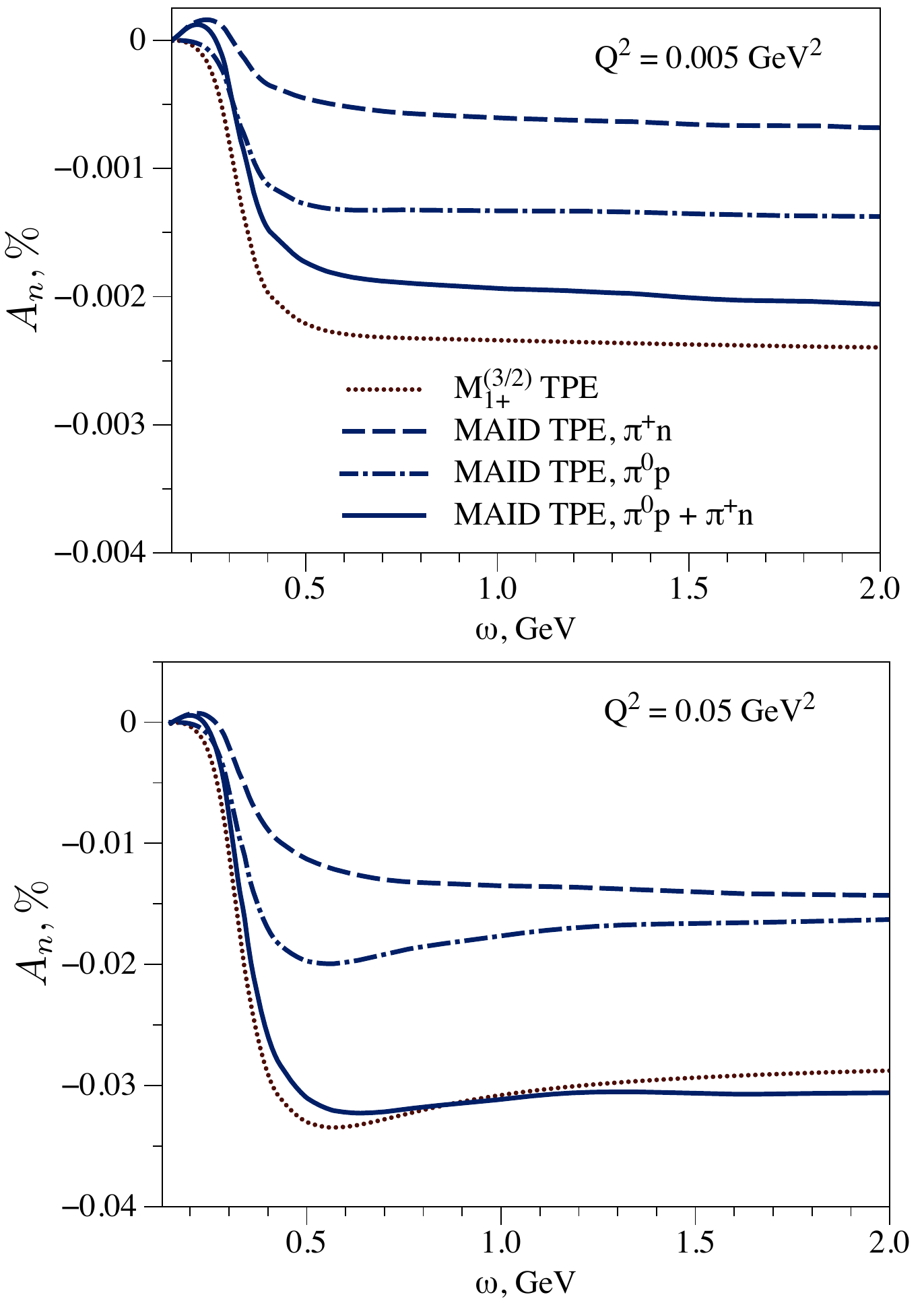}
\end{center}
\caption{The target normal spin asymmetry as a function of the electron beam energy in the laboratory frame $ \omega$ for $ Q^2 = 0.005 ~\mathrm{GeV}^2 $ (upper panel) and $ Q^2 = 0.05 ~\mathrm{GeV}^2 $ (lower panel). The inelastic contribution is calculated from the $\pi N$ intermediate states, using  MAID as input. The different curves show the contributions  from the $M^{(3/2)}_{1+}$ multipole, and  from all the  partial waves in the $ \pi^0 p$ and $ \pi^+ n$ channels separately, and that from the sum of the $ \pi^0 p$ and $ \pi^+ n $ channels (see the legend for the notations).}
\label{P33_tSSA}
\end{figure}

Exploiting Eq. (\ref{TSSA}), we check that the numerical calculations of the imaginary parts of the invariant amplitudes are in agreement with the analogous evaluation of the $ \pi N$-channel contribution to the target normal single spin asymmetry $ A_n $ of Ref. \cite{Pasquini:2004pv}. Our results are shown in Fig. \ref{P33_tSSA}.

We also check numerically that the imaginary parts of the amplitudes $ \cG_1,~\cG_2 $ vanish in the limit $ Q^2 \to 0 $ at a fixed value of $\nu$, whereas the imaginary part of the amplitude $ \cF_3 $ behaves like $ a \ln Q^2 + b $, where a and b are constants, in agreement with the low-$ Q^2 $ limit of Ref. \cite{Tomalak_PhD}.

\section{Dispersion relations}
\label{sec4}

Having specified the imaginary parts, we next evaluate the real parts of the TPE amplitudes. As the amplitudes $ \cG^{2 \gamma}_{1,2} $ are odd functions in $ \nu$, and the amplitude $ \cF^{2 \gamma}_3$ is an even function in $\nu$, they satisfy the following DRs at a fixed value of the momentum transfer $ Q^2 $ \cite{Borisyuk:2008es,Tomalak:2014sva}:
\ber \label{DRs}
 \Re {\cal{G}}^{2 \gamma}_{1,2}(\nu, Q^2) & = &  \frac{2 \nu}{\pi} \fint \limits^{~ \infty}_{\nu_{\mathrm{thr}}} \frac{\Im {\cal{G}}^{2 \gamma}_{1, 2} (\nu^\prime, Q^2)}{{\nu^\prime}^2-\nu^2}  \mathrm{d} \nu^\prime, \nonumber \\
 \Re  {\cal{F}}^{2 \gamma}_3 (\nu, Q^2) & = & \frac{2}{\pi} \fint \limits^{~ \infty}_{\nu_{\mathrm{thr}}}  \nu^\prime \frac{\Im  {\cal{F}}^{2 \gamma}_3  (\nu^\prime, Q^2)}{{\nu^\prime}^2-\nu^2}  \mathrm{d} \nu^\prime.
 \eer
The elastic contribution, corresponding with a nucleon intermediate state, has been evaluated in a previous work \cite{Tomalak:2014sva}. In this work, we evaluate the dispersive integral for the $ \pi N$ inelastic contribution, which starts from the pion production threshold $ \nu_{\mathrm{thr}} = M m_\pi + m_\pi^2 / 2 - Q^2 / 4 $. 
\begin{figure}[htp]
\begin{center}
\includegraphics[width=1.\textwidth]{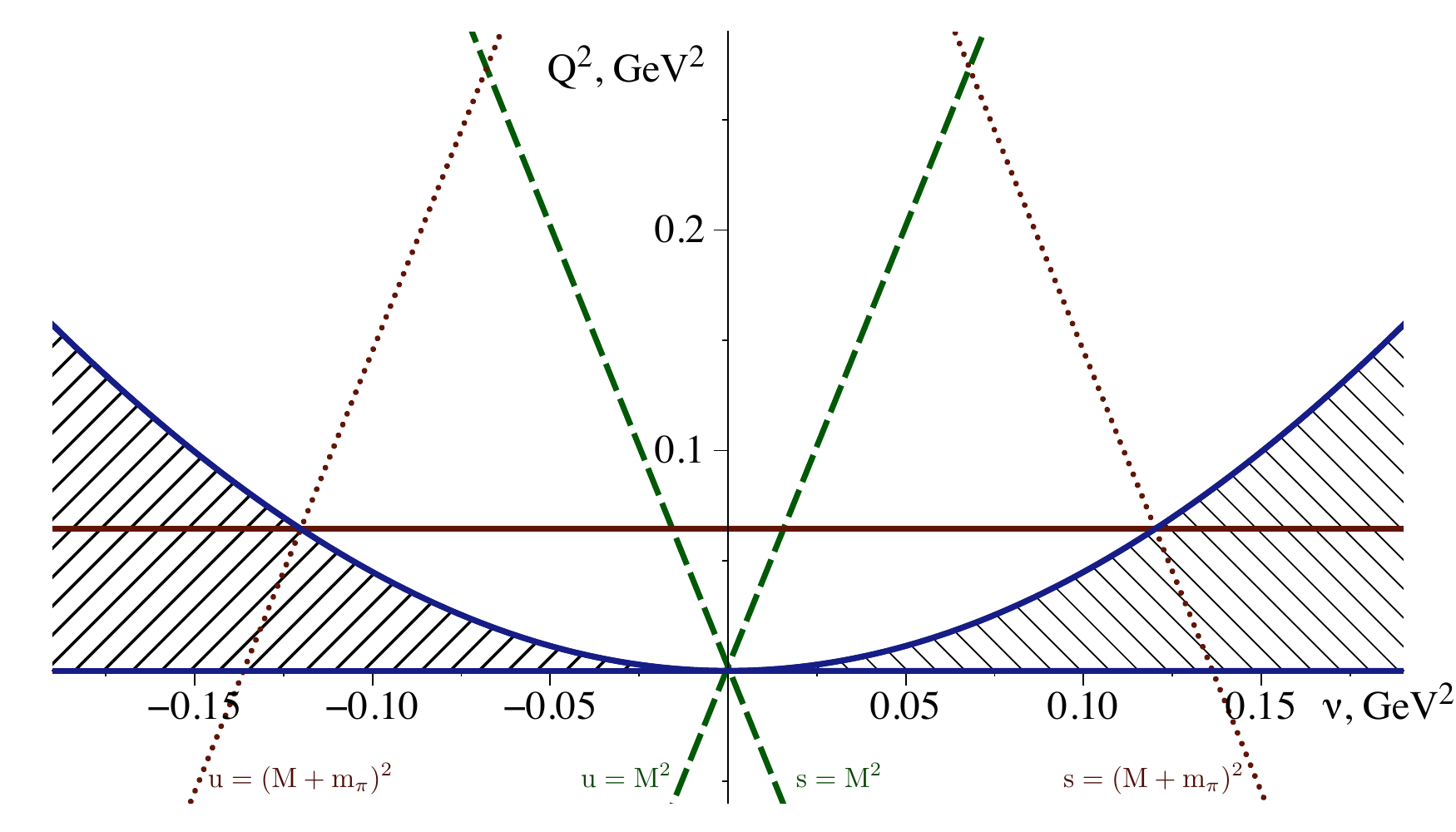}
\end{center}
\caption{Physical and unphysical regions of the kinematical variables $ \nu $ and $ Q^2 $ (Mandelstam plot) for the elastic electron-proton scattering. The hatched blue region corresponds to the physical region, the dashed green lines give the elastic threshold positions in the $s$ and $u$ channels, and the dotted red lines give the inelastic threshold positions in the $s$ and $u$ channels. The $s$-channel cuts in the TPE box diagram start at these threshold positions, yielding imaginary parts of the TPE amplitudes. The horizontal red curve at  $Q^2=0.064$ GeV$^2$ gives the boundary below which  the path of the dispersive integrals at fixed $Q^2$ runs completely in the physical region.}
\label{mandelstam_massless}
\end{figure}
 
To evaluate the dispersive integrals in Eq. (\ref{DRs}) we need to know the imaginary parts of the invariant amplitudes from the threshold energy upwards. The imaginary parts evaluated from the unitarity relations by performing a phase-space integration over physical values of the angles only yield the physical region of integration. To illustrate the physical and unphysical regions of integration,  we show in Fig. \ref{mandelstam_massless} the Mandelstam plot for the elastic electron-proton scattering. The line $ Q^2 = m^2_\pi (2 M + m_\pi)^2/(M + m_\pi)^2 \simeq 0.064 ~ \mathrm{GeV}^2 $ (indicated by the red horizontal line in Fig. \ref{mandelstam_massless}) marks a dividing line between the momentum transfers where the imaginary part due to the $ \pi N$ intermediate state in the TPE box diagram is completely contained in the physical region for the $ e^- p $ process ($ Q^2 < 0.064 ~ \mathrm{GeV}^2 $), and the region where an analytical continuation of TPE amplitudes into the unphysical region is required. In this work, we evaluate the $ \pi N $ contribution in the region where only the input from the physical region is required, i.e., for $ Q^2 < 0.064 ~ \mathrm{GeV}^2 $. We plan to cover the region  $ Q^2 > 0.064 ~ \mathrm{GeV}^2 $, where a procedure for an analytical continuation of the inelastic $ \pi N$ TPE contribution is required, in a subsequent work.
 
\begin{figure}[htp]
\begin{center}
\includegraphics[width=0.9\textwidth]{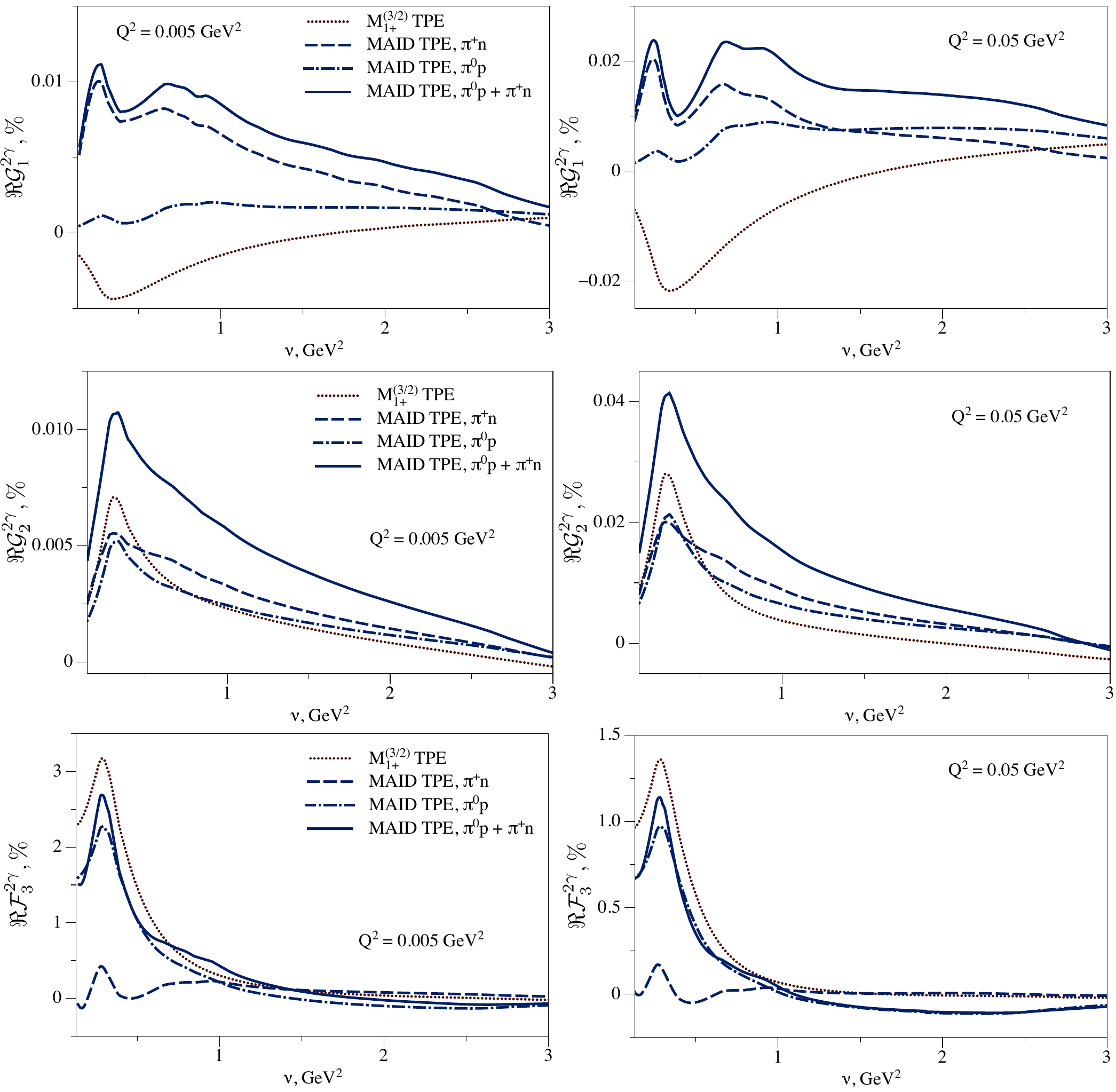}
\end{center}
\caption{Real part of the inelastic contribution to the TPE amplitudes $\cG^{2 \gamma}_{1},~\cG^{2 \gamma}_{2},~\cF^{2 \gamma}_{3}$ in the elastic electron-proton scattering for $ Q^2 = 0.005 ~\mathrm{GeV}^2 $ (left panels) and $ Q^2 = 0.05 ~\mathrm{GeV}^2 $ (right panels). 
The different curves show the results obtained with the input from MAID with only the $M^{(3/2)}_{1+}$ multipole, those obtained with all $ \pi N$ partial waves in the $ \pi^+ n$, 
and $ \pi^0 p$ channels separately, and those found from the sum of the $ \pi^0 p$ and $ \pi^+ n$ channels (see the legend for the notations).
 The upper integration limit in the DR was chosen as $\nu_{\mathrm{max}} = 12 ~\mathrm{GeV}^2 $. 
}
\label{P33_real_G1_G2_F3}
\end{figure}

We present the real parts of the TPE amplitudes $ \cG^{2 \gamma}_{1},~\cG^{2 \gamma}_{2},~\cF^{2 \gamma}_{3}$ for $ Q^2 = 0.005 ~\mathrm{GeV}^2 $ and $ Q^2 = 0.05 ~\mathrm{GeV}^2 $ in Fig. \ref{P33_real_G1_G2_F3}. We show results for the $M^{(3/2)}_{1+}$ multipole as well as for all $ \pi N$ partial waves using MAID as input. We integrate the dispersion relations up to $\nu_{\mathrm{max}} = 12 ~\mathrm{GeV}^2 $. 

Similarly to the behavior of the imaginary parts, the real parts of the amplitudes $ \cG_1,~\cG_2 $ vanish in the limit $ Q^2 \to 0 $ at a fixed value of $\nu$, and the real part of the amplitude $ \cF_3 $ behaves like $ a \ln Q^2 + b $, where a and b are constants, in agreement with the low-$ Q^2 $ limit of Ref. \cite{Tomalak_PhD}.

\section{Results and discussion}
\label{sec5}

\begin{figure}[h]
\begin{center}
\includegraphics[width=0.58\textwidth]{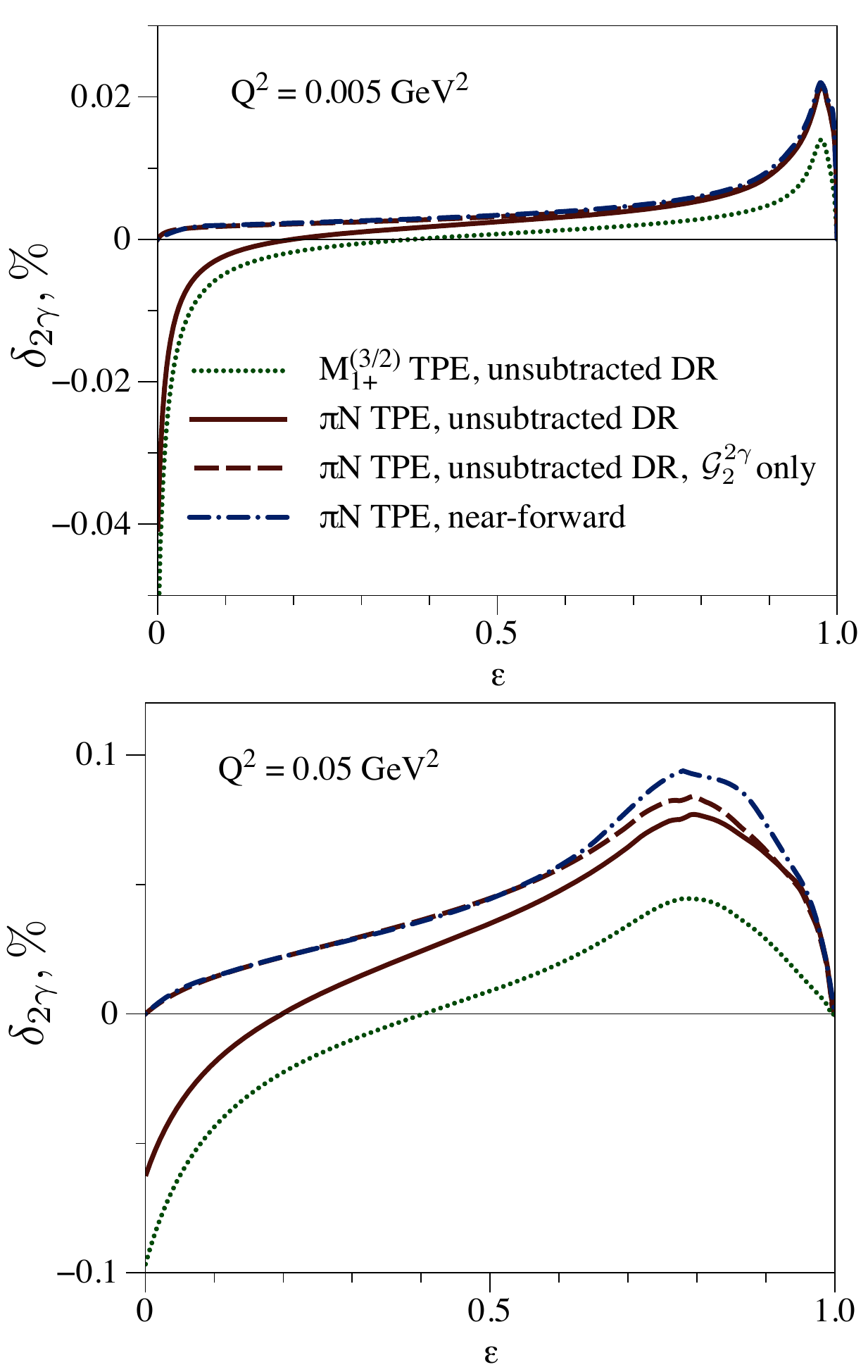}
\end{center}
\caption{TPE correction within the unsubtracted DRs using all $\pi N$ partial waves from MAID, compared with the contribution from the amplitude $ \cG_2^{2 \gamma}$ only, as well as with the {\it near-forward} approximation and the $ M^{(3/2)}_{1+}$ multipole contributions for $ Q^2 = 0.005 ~\mathrm{GeV}^2 $ (upper panel) and $ Q^2 = 0.05 ~\mathrm{GeV}^2 $ (lower panel).}
\label{all_partial_waves_unsubtracted_DR_vs_NF}
\end{figure}

Substituting the real parts of the TPE amplitudes into Eq. (\ref{unpolarized_cross_section}), we obtain the unsubtracted DR result for the inelastic TPE correction to the electron-proton scattering cross section. In Fig. \ref{all_partial_waves_unsubtracted_DR_vs_NF}, we compare  the result for the TPE correction from all $ \pi N$ partial waves in MAID with the TPE correction due to the $ \cG_2^{2 \gamma}$ amplitude only, which is the leading term in Eq. (\ref{unpolarized_cross_section}) for the scattering at small momentum transfer. As an independent check, we also show the evaluation of the TPE corrections in the {\it near-forward} approximation of Ref. \cite{Tomalak:2015aoa} obtained with the same experimental input from MAID. 
We restrict the upper limit of integration to $W_{\mathrm{MAID}} = 2~\mathrm{GeV}$. We first notice a good agreement over the whole $\varepsilon$ range between the {\it near-forward} calculation and the DR evaluation using the contribution only from the amplitude $ \cG_2^{2 \gamma}$. As a matter of fact, the {\it near-forward} calculation, corresponding to $ \varepsilon \to 1$ in Eq. (\ref{unpolarized_cross_section}), is dominated by the $\cG_2^{2 \gamma} $ term at small
momentum transfer. In the large-$ \varepsilon $ range, these estimates are also in good agreement  with the full DR  calculation, which includes the contributions from all the electron-proton scattering amplitudes $ \cG_1,~\cG_2 $ and $ \cF_3 $. We estimate the region of applicability of the {\it near-forward} approximation by a simple criterion that the DR result $ \delta_{2 \gamma}^{\mathrm{DR}}$ is reproduced by the {\it near-forward} calculation $ \delta_{2 \gamma}^{\mathrm{NF}}$ within 15~\%: $ |1 - \delta_{2 \gamma}^{\mathrm{DR}} / \delta_{2 \gamma}^{\mathrm{NF}} | < 0.15$. Consequently, the approximation can be used for $ \varepsilon \gtrsim 0.7 $ at $ Q^2 = 0.005~\mathrm{GeV}^2 $ and for $ \varepsilon \gtrsim 0.9 $ at $ Q^2 = 0.05~\mathrm{GeV}^2 $.

\begin{figure}[h]
\begin{center}
\includegraphics[width=0.62\textwidth]{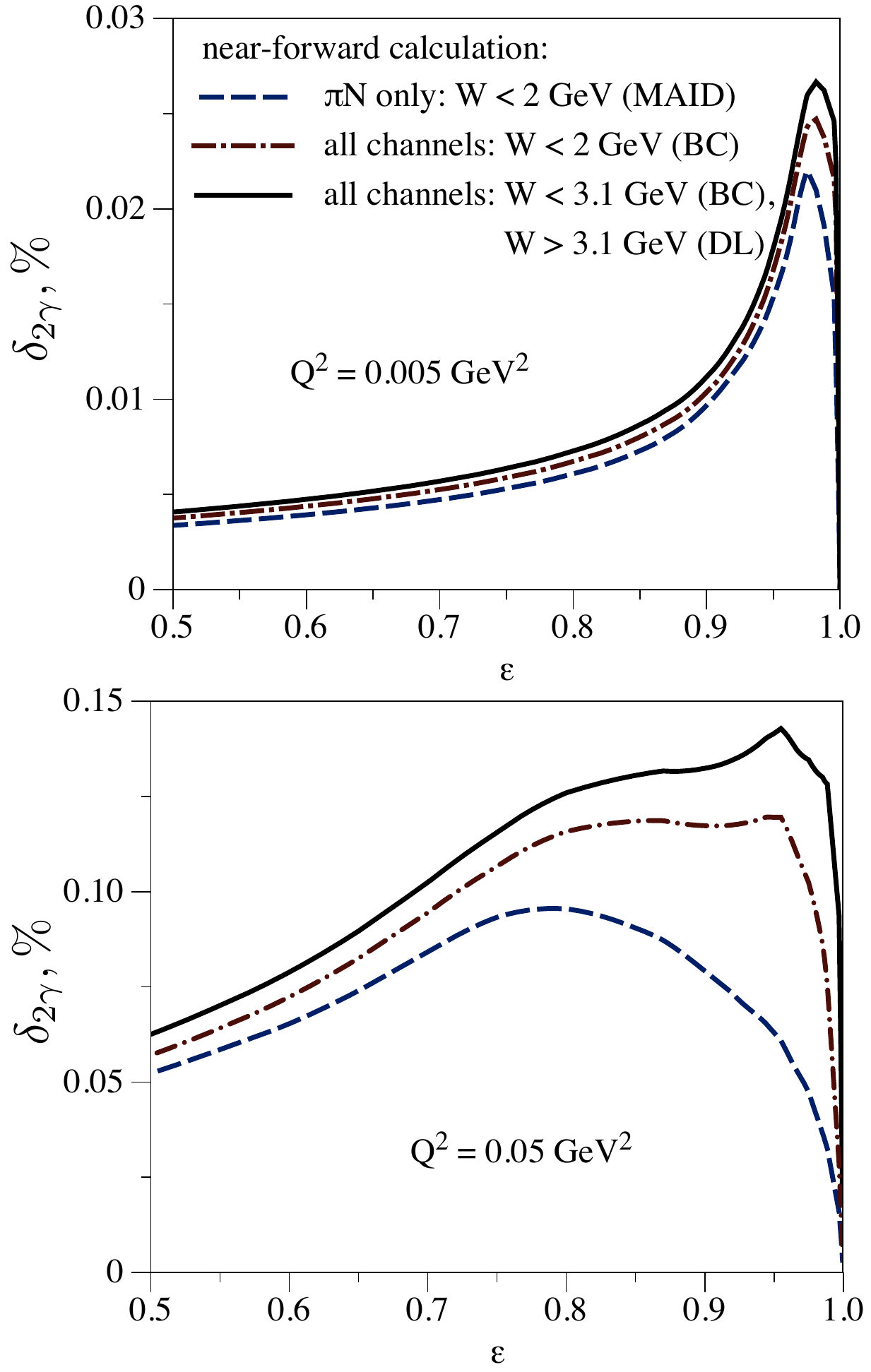}
\end{center}
\caption{Results for the inelastic TPE correction in the {\it near-forward} approximation of Ref. \cite{Tomalak:2015aoa} for $ Q^2 = 0.005 ~\mathrm{GeV}^2 $ (upper panel) and $ Q^2 = 0.05 ~\mathrm{GeV}^2 $ (lower panel). The $ \pi N $ TPE correction with input from MAID is compared with the total inelastic TPE based on the BC fit \cite{Christy:2007ve}, as well as the BC fit  combined with the DL fit \cite{Donnachie:2004pi}.}
\label{P33_unsubtracted_DR_low_Q2_all}
\end{figure}

As the $ \pi N$ intermediate-state contribution is expected to dominate the inelastic TPE correction at low momentum transfer, we can quantify the contributions beyond $ \pi N$ by performing different evaluations in the {\it near-forward} approximation of Ref. \cite{Tomalak:2015aoa}. As the latter estimate is based on the input of the full proton structure functions, a comparison within the same {\it near-forward} evaluation but based on the structure functions 
calculated from the single $ \pi N $ channel in MAID allows one to get an estimate of contributions beyond the $ \pi N $ channel (e.g. $ \pi \pi N, ...$).  In Fig. \ref{P33_unsubtracted_DR_low_Q2_all}, we show the resulting inelastic TPE correction evaluated in the {\it near-forward} approximation based on the structure function input from the fit performed by Christy and Bosted (BC) \cite{Christy:2007ve} (valid for $ W < 3.1~\mathrm{GeV} $) supplemented with the Donnachie-Landshoff (DL) fit \cite{Donnachie:2004pi} (for $ W > 3.1~\mathrm{GeV} $). We compare this result with the inelastic TPE in the {\it near-forward} approximation using the $ \pi N $ contribution from MAID (valid for $ W < 2~\mathrm{GeV} $). For $ Q^2 = 0.005 ~\mathrm{GeV}^2 $, one notices that the $ \pi N $ channel completely dominates the inelastic TPE correction. Already at the momentum transfer  $Q^2 = 0.05 ~\mathrm{GeV}^2 $, the contributions of higher channels ($ \pi \pi N, ...$) become relevant.

\begin{figure}[H]
\begin{center}
\includegraphics[width=0.75\textwidth]{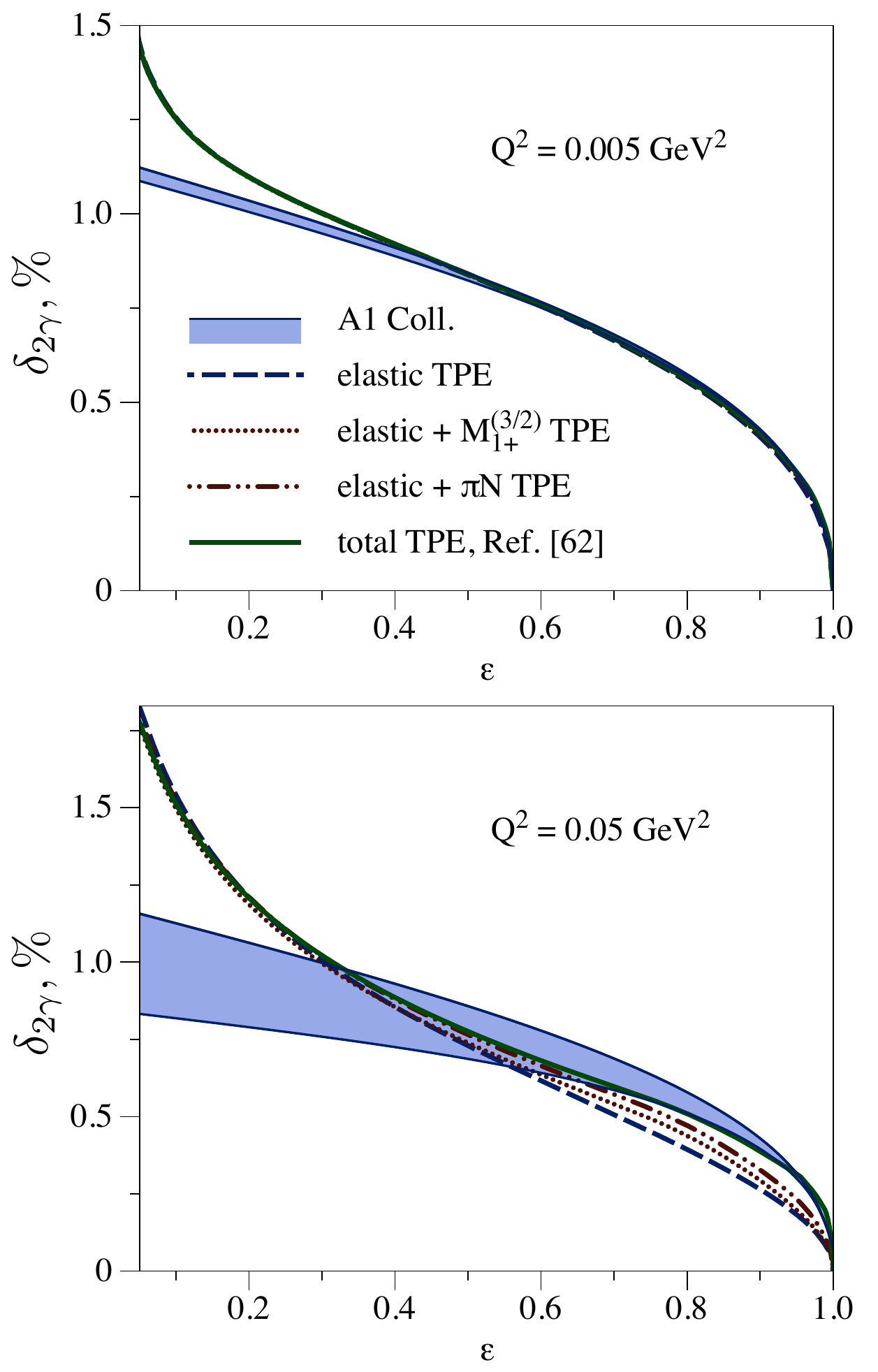}
\end{center}
\caption{Unsubtracted DR evaluation of the different TPE corrections: elastic TPE, elastic + $M^{(3/2)}_{1+}$  TPE, elastic + $ \pi N $ TPE, and total inelastic TPE in the {\it near-forward} approximation from Ref. \cite{Tomalak:2015aoa} for $ Q^2 = 0.005 ~\mathrm{GeV}^2 $ (upper panel) and $ Q^2 = 0.05 ~\mathrm{GeV}^2 $ (lower panel). The blue bands show the empirical TPE extraction from Ref. \cite{Bernauer:2013tpr}.
}
\label{P33_unsubtracted_DR}
\end{figure}

In Fig. \ref{P33_unsubtracted_DR}, we compare the sum of the elastic and $ \pi N $ contributions (with the upper limit of integration  $ \nu_{\mathrm{max}} = 12~\mathrm{GeV}^2 $) using the unsubtracted DR framework with the total TPE of Ref. \cite{Tomalak:2015aoa}  in the {\it near-forward} approximation. We also show the elastic TPE and the empirical TPE fit of Ref. \cite{Bernauer:2013tpr}. For the very small $ Q^2 = 0.005~\mathrm{GeV^2}$ value, the inelastic TPE correction is much smaller than the elastic TPE. At $ Q^2 = 0.005~\mathrm{GeV^2}$ the Feshbach correction \cite{McKinley:1948zz} is a very good approximation for the scattering at small angles. For scattering at backward angles, a precise estimate of the TPE correction mainly requires us to accurately account for the elastic electromagnetic proton structure as encoded in the form factors. At the larger $ Q^2 = 0.05~\mathrm{GeV^2}$ value, one notices that in the intermediate- and large-$\varepsilon$ regions, the inelastic TPE correction is necessary to describe the empirical extraction.


\begin{figure}[h]
\begin{center}
\includegraphics[width=0.48\textwidth]{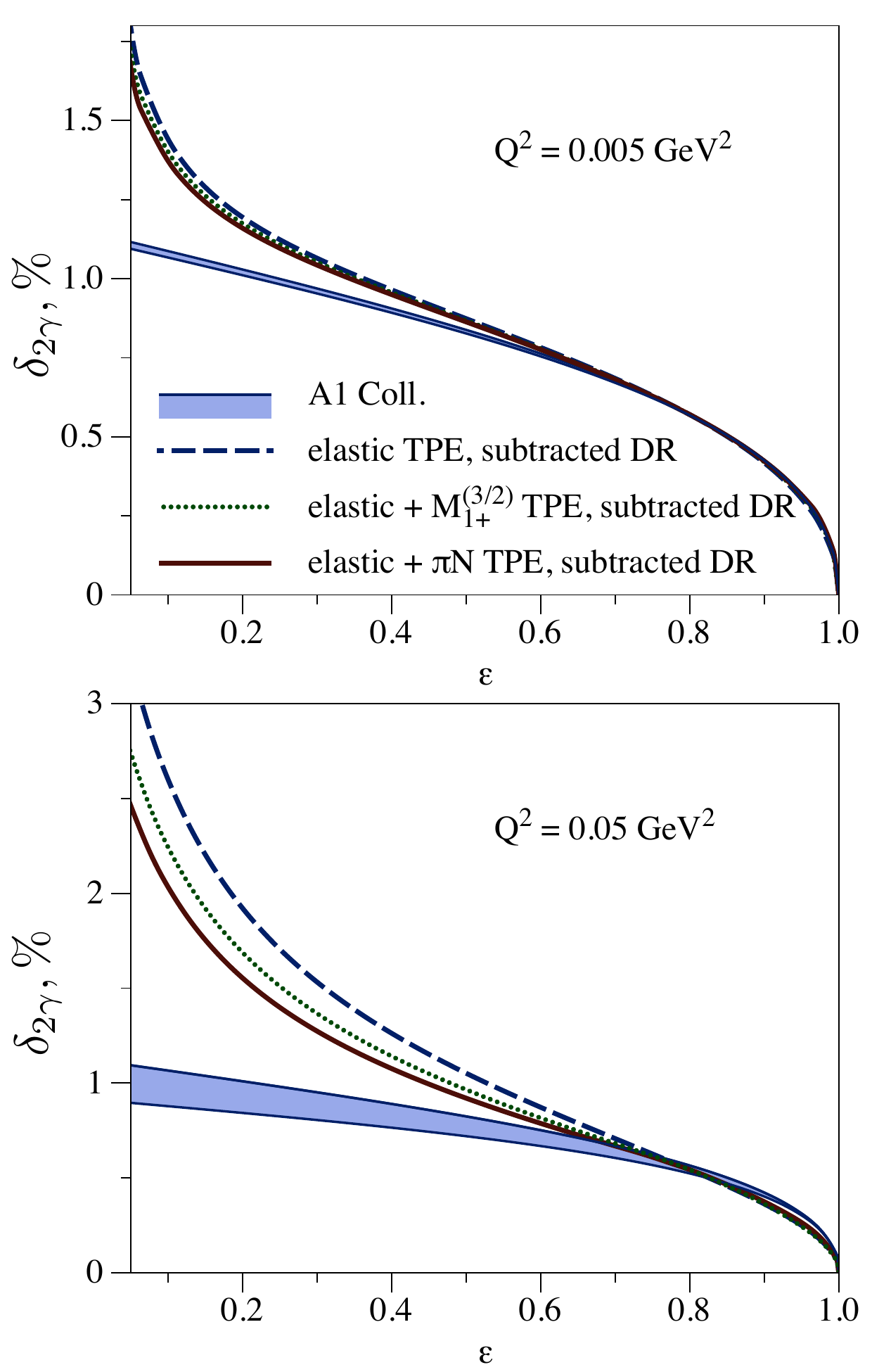}
\end{center}
\caption{Subtracted DR-based prediction for the elastic TPE \cite{Tomalak:2014sva} and the elastic + inelastic TPE, where for the inelastic TPE all $ \pi N$ channels in MAID are accounted for. Upper panel: $ Q^2 = 0.005 ~\mathrm{GeV}^2 $. Lower panel: $ Q^2 = 0.05 ~\mathrm{GeV}^2 $. The subtracted DR curves correspond to the subtraction point: $ \varepsilon_0 = 0.8 $. The blue bands show the empirical TPE extraction from Ref. \cite{Bernauer:2013tpr}.}
\label{P33_subtracted_DR}
\end{figure}

In order to further reduce the model uncertainty and the sensitivity to inelastic contributions at higher energies, we use the subtracted DR formalism of Ref. \cite{Tomalak:2014sva} with the subtraction in the amplitude $ \cF_3^{2 \gamma}$ and show our results in Fig. \ref{P33_subtracted_DR}. Our results are within the band of the empirical TPE extraction for $ \varepsilon \gtrsim 0.6$. Note, however, that the empirical parametrization is extrapolated into the region where data are not available; i.e., $ \varepsilon \lesssim 0.3$ for $ Q^2 = 0.005~\mathrm{GeV}^2$ and $ \varepsilon \lesssim 0.45$ for $ Q^2 = 0.05~\mathrm{GeV}^2$.   Therefore, the discrepancies between the predictions and empirical fits could be due to the missing contribution of higher-energy intermediate states ($ \pi \pi N, ...$) in the theoretical calculation and/or the extrapolation of the fits outside the kinematics covered by the measurements.

\section{Conclusions and Outlook}
\label{sec6}

In this work, we have extended a previously developed dispersive framework for the TPE correction to the elastic electron-proton scattering by including the contribution of all $ \pi N $ intermediate states. The imaginary parts of TPE amplitudes were reconstructed by unitarity relations in terms of physical (on-shell) electroproduction amplitudes \cite{Drechsel:2007if}. To provide the most realistic evaluation of these imaginary parts, we use the state-of-the-art MAID 2007 parametrization for the pion electroproduction amplitudes as input in our calculations. Within unsubtracted DRs, the real parts of TPE amplitudes were reconstructed in the kinematical region $ Q^2 < 0.064~\mathrm{GeV}^2$, where the analytical continuation of imaginary parts into the unphysical region of the TPE scattering amplitudes is not required. Exploiting these real parts, the TPE correction to the unpolarized cross section was evaluated, providing first quantitative estimates of the full contribution of the $ \pi N $ channel to the inelastic TPE correction.

At low momentum transfer and small scattering angles, the unsubtracted DR result was found to be in a good agreement with the {\it near-forward} TPE calculation of Ref. \cite{Tomalak:2015aoa} based on the same experimental input. This comparison provides the region of applicability of the {\it near-forward} approximation. Our results for the kinematics of the MAMI experiments \cite{Bernauer:2013tpr} show that the $ \pi N$ contribution has the same sign as the Feshbach correction \cite{McKinley:1948zz} for the scattering in the forward direction and the opposite sign in the backward scattering, where we found a cancellation between the $M^{(3/2)}_{1+}$ multipole due to the $\Delta$(1232) resonance and contributions due to nonresonant $ \pi N$ production as well as higher resonance states. We also applied the subtracted DR formalism extending a previous work  \cite{Tomalak:2014sva} with the inclusion of the $\pi N$ inelastic contribution. In this way, the uncertainties due to neglecting higher intermediate states, such as $\pi \pi N$, are reduced.While our unsubtracted DR calculations of the $ \pi N$ contribution improve the agreement with the empirical TPE extractions at low $Q^2$ and forward angles, they are much smaller than the leading elastic contribution at backward angles. The latter indicates a significant departure at backward angles from the simplified (linear) $\varepsilon$ parametrization used in the current phenomenological fits. This calls for a reanalysis of backward angle data at low $Q^2$, which is of relevance in a precise extraction of the proton magnetic radius and the proton magnetic form factor at low $Q^2$ values. Such application and reanalysis is planned in a future work. 

In a subsequent paper, we also plan to present the extension of our work to larger values of $Q^2$. For $Q^2 > 0.064~\mathrm{GeV}^2$, we will outline a procedure how to analytically continue the TPE amplitudes outside the physical region based on phenomenological $ \pi N$ input.

\section*{Acknowledgements}

We thank Lothar Tiator for useful discussions and providing us with MAID programs, Dalibor Djukanovic for providing us with the access to computer resources, Volodymyr Schubny for discussions about the narrow-$\Delta$ contribution, and Slava Tsaran for his script for an online access of MAID. This work was supported in part by the Deutsche Forschungsgemeinschaft DFG in part through the Collaborative Research Center [The Low-Energy Frontier of the Standard Model (SFB 1044)], in part through the Graduate School [Symmetry Breaking in Fundamental Interactions (DFG/GRK 1581)], and in part through the Cluster of Excellence [Precision Physics, Fundamental Interactions and Structure of Matter (PRISMA)].

\appendix

\section{Kinematics of pion production in $ep$ scattering}
\label{app1}

\begin{figure}[h]
\begin{center}
\includegraphics[width=.45\textwidth]{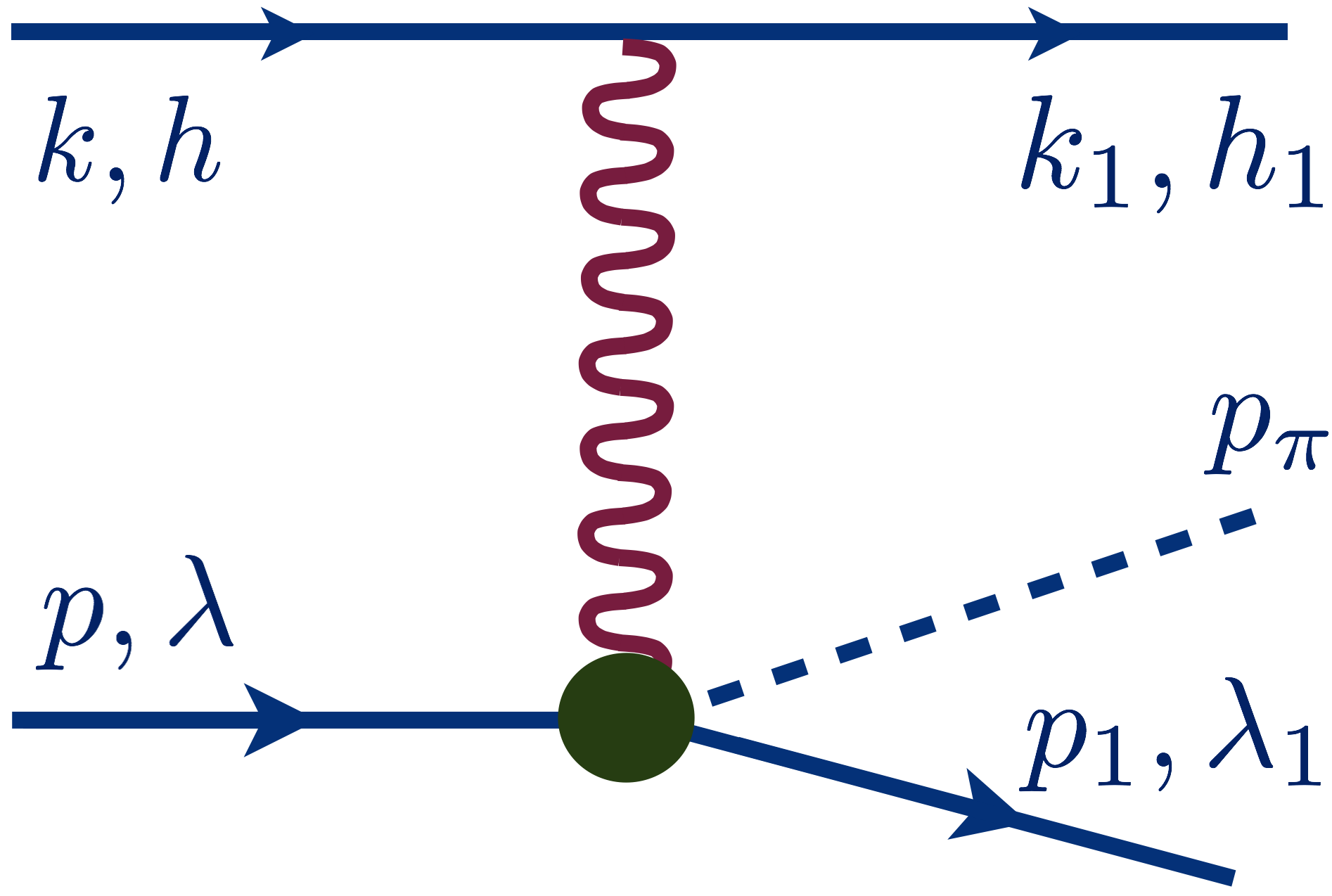}
\end{center}
\caption{Pion production in the electron-proton scattering.}
\label{piN_electroproduction}
\end{figure}

In this work, we study the pion-nucleon intermediate-state TPE correction exploiting the on-shell information of pion electroproduction. The production of pions in the scattering of electrons off a proton target, $ e(k, h) + p (p, \lambda) \to e(k_1, h_1) + p (p_1, \lambda_1) + \pi^0 (p_\pi)$ or $ e(k, h) + p (p, \lambda) \to e(k_1, h_1) + n (p_1, \lambda_1) + \pi^+ (p_\pi)$, see Fig. \ref{piN_electroproduction}, is completely described by five kinematical variables. We conventionally use the squared energy $ s = ( p + k )^2 $, the squared momentum transfer between the electrons $ Q^{2}_{1} = - q^{2}_{1} = - (k - k_1)^2 $,  the momentum transfer variable between the nucleons $ t_\pi = (p_1 - p)^2 $, the squared invariant mass of the pion-nucleon system $ W^2 = (p_1 +p_\pi )^2 = (p+q_1)^2$, and one relative angle between the electron and hadron production planes.

We study here the kinematics of pion production in the electron-proton c.m. reference frame, as we choose this frame to relate the $e p$ helicity amplitudes to the invariant amplitudes in Sec. \ref{sec2}. The initial (final) electron energy $ \omega_{\mathrm{cm}} $ ($ \omega_1 $) and the momentum $ | \vec{k}_{\mathrm{cm}} | $ ($ | \vec{k}_1 | $)  are given by 
\ber
 \omega_{\mathrm{cm}} =  | {\vec{k}}_{\mathrm{cm}} | = \frac{s - M^2}{2\sqrt{s}}, \qquad \omega_1 =   | {\vec{k}}_1 |  = \frac{s - W^2 }{2\sqrt{s}}.
\eer

The initial ($k$) and final ($k_1$) electron four-momenta (neglecting the $ e$ mass) are given by
\ber \label{kinematics_piN}
 k & = & (\omega_{\mathrm{cm}}, 0, 0, \omega_{\mathrm{cm}} ), \\
 k_1 & = & (\omega_1  ,\omega_1 \sin \theta_1 \cos \phi_1 ,\omega_1 \sin \theta_1 \sin \phi_1 , \omega_1\cos \theta_1),  
\eer 
with the electron scattering angles $ \theta_1 $ and $ \phi_1 $. 

The pion four-momentum $ p_\pi = \left(E_\pi, \vec{p}_\pi \right) $ can be expressed in terms of the pion angles $ \theta_\pi,~\phi_\pi $ as
\ber
 p_\pi & = &  \left(\sqrt{\vec{p}_\pi^{\,2}+m^2_\pi}, |\vec{p}_\pi| \sin \theta_\pi \cos \phi_\pi, |\vec{p}_\pi| \sin \theta_\pi \sin \phi_\pi, |\vec{p}_\pi| \cos \theta_\pi \right).
\eer
The relative angle between the final electron and pion, denoted by $ \Theta_\pi $, is given by
\ber
\cos \Theta_\pi  & \equiv & \cos \left( \hat{k}_1 \cdot \hat{p}_\pi \right)=  \sin \theta_1  \sin \theta_\pi \cos \left( \phi_1 - \phi_\pi \right) + \cos \theta_1  \cos \theta_\pi.
\eer

The initial ($ p $) and final ($ p_1 $) nucleon four-momenta are given by
\ber
p  = \left(\frac{s+M^2}{2\sqrt{s}}, 0, 0, -\omega_{\mathrm{cm}} \right), \qquad
p_1 = \left( \sqrt{(\vec{k}_1 + \vec{p}_\pi)^{2} + M^2}, - \vec{k}_1 - \vec{p}_\pi \right).
\eer

We first analyze the special cases in the phase space of the final particles. The minimum of the final electron momentum $  | {\vec{k}}_1 | = 0 $ corresponds to the geometrical configuration, when the pion and nucleon are moving in opposite directions, i.e.
\ber
  E_\pi + E_1 =  \sqrt{s}, \qquad
|\vec{p}_\pi|  =  |\vec{p}_1| = \frac{\sqrt{\Sigma \left( s, M^2, m_\pi^2 \right)}}{2 \sqrt{s}},
\eer
with the kinematical triangle function $  \Sigma (s, ~M^2, ~m_\pi^2) \equiv \left[ s-(M +m_\pi)^2 \right] \left[s-(M -m_\pi \right)^2] $. The maximum $k_{1}^{\mathrm{max}}$ of $ | \vec{k}_1 | $ corresponds to the minimum of $W^2 = (M+m_\pi)^2 $. It is reached for the case of the same pion and nucleon momenta directions, which are opposite to the electron momentum direction, i.e.,  $ \cos \Theta_\pi  = - 1 $. The corresponding particle momenta are given by
\ber
  k^{\mathrm{max}}_1 &=& \frac{s - (M + m_\pi)^2}{2 \sqrt{s}}, \\
p_1^{\mathrm{max}}& = &  \frac{M}{M + m_\pi} k^{\mathrm{max}}_1 ,\quad 
p_\pi^{\mathrm{max}} = \frac{m_\pi}{M + m_\pi} k^{\mathrm{max}}_1.
\eer

Accounting for the energy conservation, we obtain the following expression for the pion momentum:
\ber \label{pion_momentum}
p_\pi^\pm \equiv - \left( \frac{ W^2 - M^2 + m^2_\pi }{W^2 + \omega_1^2 \sin^2 \Theta_\pi} \right) \frac{\omega_1 \cos \Theta_\pi}{2} \pm \frac{ \sqrt{\Sigma(W^2, M^2, m_\pi^2)- 4 m^2_\pi \omega_1^2 \sin^2 \Theta_\pi }}{W^2 + \omega_1^2 \sin^2 \Theta_\pi} \frac{\sqrt{s} -  \omega_1}{2}. \nonumber \\
\eer
In the kinematical region $ \Sigma(W^2, M^2, m_\pi^2) \ge 4 m^2_\pi \omega_1^2 $, only the  solution $ p_\pi^+ $ is positive. In the region $ \Sigma(W^2, M^2, m_\pi^2) \le 4 m^2_\pi \omega_1^2 $, both solutions are positive and can be realized. In this case, the pion can be scattered only into the backward cone with respect to the electron momentum direction, $ \Theta_\pi >  \Theta^0_\pi > \frac{\pi}{2} $, with the limiting half-angle of the cone $ \Theta^0_\pi $ given by
\ber
\sin^2 \Theta^0_\pi = \frac{\Sigma(W^2, M^2, m_\pi^2)}{4 m^2_\pi \omega_1^2}.
\eer
The two distinct geometrical configurations  can be obtained only for the invariant mass 
in the range $[W_{\mathrm{min}}, W_0]$, with $W_{\mathrm{min}}$
 corresponding to the point where the positive and negative solutions  coincide, i.e.
\ber
W^2_{\mathrm{min}} = \frac{ M^2 + m^2_\pi \cos^2 \Theta_\pi }{ s - m^2_\pi \sin^2 \Theta_\pi } s + \frac{ 2 M m_\pi s}{  s - m^2_\pi \sin^2 \Theta_\pi } \sqrt{1 + \frac{\Sigma \left( s, M^2, m^2_\pi \right)}{4M^2 s} \sin^2 \Theta_\pi},
\eer
and $W_0$ corresponding to $ \sin^2 \Theta^0_\pi = 1$, i.e.
\ber
W^2_0 & = & \frac{ s + M^2 - m_\pi^2 }{s - m_\pi^2} m_\pi \sqrt{s} + \frac{ s M^2 }{s - m_\pi^2}  \underset{ \sqrt{s} \ge M + m_\pi}{ \ge } (M+m_\pi)^2,
\eer
where the minimum value  of $W_0$ is at the pion production threshold for $s = ( M +m_\pi)^2.$
\begin{figure}[h]
\begin{center}
\includegraphics[width=1.\textwidth]{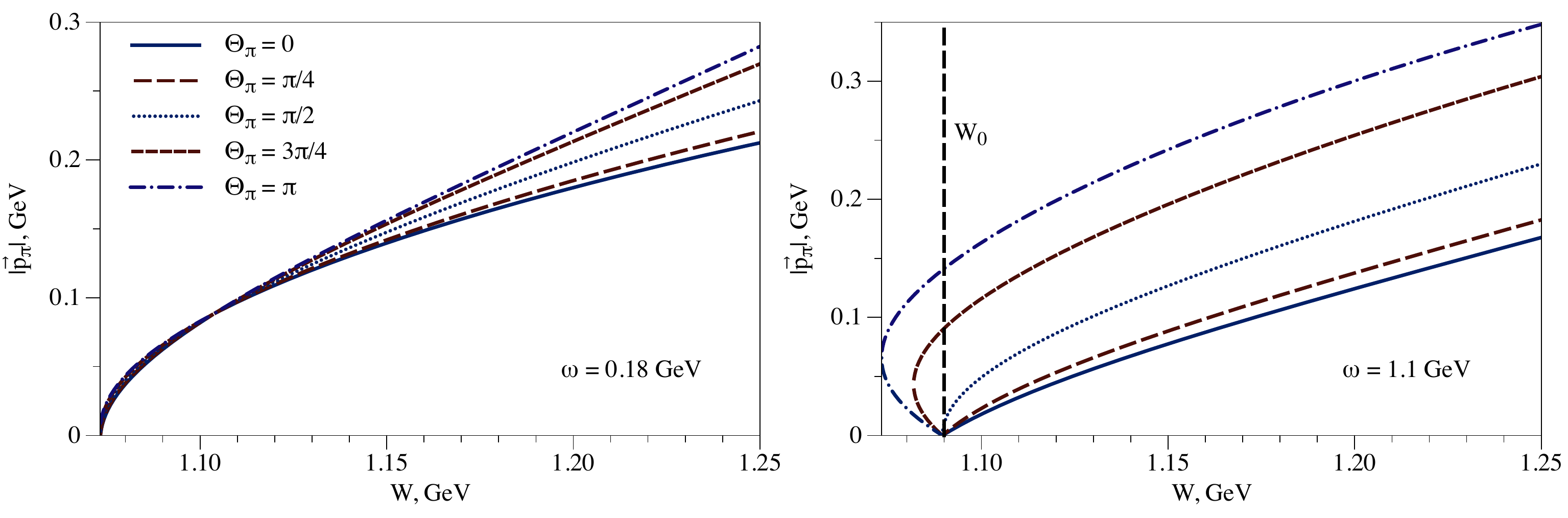}
\end{center}
\caption{The pion momentum (in the $ e  p $ c.m. frame) as a function of the invariant mass $ W $ of the $ \pi^0 p $ state in the pion electroproduction process with the electron beam energy $ \omega = 0.18~\mathrm{GeV} $ (left panel) and $ \omega = 1.1~\mathrm{GeV} $ (right panel). The vertical dashed line at fixed $ W=W_0 $ indicates the value of the invariant mass  below which there exist two distinct solutions for the pion momentum.}
\label{kinematics_p_pi}
\end{figure}

We show the dependence of $ |\vec{p}_\pi| $ on the invariant mass $ W $ of the pion-nucleon state ($ \pi^0 p $) for the typical kinematics of experiments in Fig. \ref{kinematics_p_pi}. At the lowest MAMI beam energy ($ \omega = 0.18~\mathrm{GeV}$), $W_0  \approx 1.07331 ~\mathrm{GeV}$ is slightly above the pion production threshold $W_{\mathrm{thr}} \approx 1.07325 ~\mathrm{GeV}$, and  the contribution of the region with two solutions is negligible, while at larger beam energies, both solutions should be accounted for.

\section{Invariant amplitudes for pion electroproduction}
\label{app2}

The dispersive evaluation of the $ \pi N$-channel TPE contribution to the real parts of the
electron-proton invariant amplitudes (see Fig. \ref{piN_TPE}) requires the pion production amplitudes
 in the one-photon exchange approximation as input (see Fig. \ref{unitarity_piN}). 
 Moreover, the experimental extraction of the pion electroproduction amplitudes is 
 performed in the approximation of one-photon exchange. 
 The conventional Lorentz-invariant expression for the pion electroproduction amplitude $  T^{1 \gamma}_{\pi N} $ is given by
\ber \label{piN_1ph_amplitude}
 T^{1 \gamma}_{\pi N} & = & - \frac{e^2}{Q^{2}_{1}} \bar{u} (k_1,h_1) \gamma_\mu u (k,h) \cdot \bar{N}(p_1,\lambda_1) J^\mu_{\pi N}  \left(p_\pi, p_1, p\right)  N(p,\lambda),
\eer
with  the kinematics of the process  described in Appendix \ref{app1}.
Accounting for gauge invariance as well as  parity conservation, the pion electroproduction  current can be parametrized in terms of six invariant amplitudes  $A_i$~\cite{Dennery:1961zz,Berends:1967vi,Pasquini:2007fw}: \footnote{Note that our definition of the invariant amplitudes $A_i$
\cite{Pasquini:2004pv} differs from the MAID paper \cite{Pasquini:2007fw},
where the electric charge $e$ is included in the amplitudes.}

\ber
J^\mu_{\pi N} \left(p_\pi, p_1, p\right) & = & \sum\limits^6_{i=1} A_i \left( W^2 , t_\pi, Q^{2}_{1} \right) M^\mu_i, \\
M^\mu_1&=&
-\frac{1}{2}i\gamma_5\left(\gamma^\mu \gamma\cdot q_1 - \gamma \cdot q_1 \,\gamma^\mu\right)\, ,
\\
M^\mu_2&=&2i\gamma_5\left(P^\mu\, q_1 \cdot(p_\pi-\frac{1}{2} q_1)-
(p_\pi-\frac{1}{2} q_1)^\mu\, q_1 \cdot P\right)\, ,\\
M^\mu_3&=&-i\gamma_5\left(\gamma^\mu\, q_1 \cdot p_\pi
- \gamma\cdot q_1\, p_\pi^\mu\right)\, ,\label{eq:M1-6}  \\
M^\mu_4&=&-2i\gamma_5\left(\gamma^\mu\, q_1 \cdot P
- \gamma\cdot q_1 \, P^\mu\right)-2 M \, M^\mu_1\, ,\\
M^\mu_5&=&i\gamma_5\left(q_1^\mu\, q_1 \cdot p_\pi
+ Q^{2}_{1} p_\pi^\mu\right)\, ,\\
M^\mu_6&=&-i\gamma_5\left(\gamma \cdot q_1\, q_1^{\mu}+ Q^{2}_{1} \gamma^\mu\right)\, ,
\end{eqnarray}
with $P=(p + p_1)/2$. This expression is manifestly gauge invariant; i.e., each covariant $ M^\mu_i $ satisfies $ q_{1\mu} M^\mu_i = 0 $. For the numerical implementation we exploit the invariant amplitudes $ A_i $ from the MAID fit (version 2007) \cite{Drechsel:1998hk,Drechsel:2007if}. 

We exploit also the conjugated amplitude for the second photon in Fig. \ref{unitarity_piN}. The nucleon current enters it in the complex conjugated form
\ber
\left( \bar{N}(p_1, \lambda_1) J^\mu_{\pi N}  \left(p_\pi, p_1, p\right)   N(p, \lambda) \right)^* & = &  \bar{N}(p, \lambda) \tilde{J}^{\mu}_{\pi N}  \left(p_\pi, p_1, p\right)  N(p_1, \lambda_1),
\eer
with the conjugated pion production current
\ber
 \tilde{J}^{ \mu}_{\pi N}  \left(p_\pi, p_1, p\right)  = \sum\limits^6_{i=1} A_i^* \left( W^2, t_\pi, Q^{2}_{1} \right) \tilde{M}^\mu_i,
\eer
and covariants
\begin{eqnarray} \label{structures_invariant}
\tilde{M}^\mu_1 = - M^\mu_1\, , \qquad
\tilde{M}^\mu_2 = M^\mu_2\, , \qquad
\tilde{M}^\mu_3 = - M^\mu_3\, ,\label{eq:M1-6} \nonumber \\
\tilde{M}^\mu_4 = - M^\mu_4\, ,\qquad
\tilde{M}^\mu_5 = M^\mu_5\, , \qquad
\tilde{M}^\mu_6 = - M^\mu_6\ . 
\end{eqnarray}
 
\appendix


\begin{thebibliography}{99}

\bibitem{hofs53}
R. Hofstadter, H.R. Fechter and J.A. McIntyre,
Phys. Rev.  92 (1953) 978.

\bibitem{hofs55}
R. Hofstadter and R.W. McAllister, Phys. Rev. 98 (1955) 217.

\bibitem{Jones:1999rz} 
  M.~K.~Jones {\it et al.} [Jefferson Lab Hall A Collaboration],
  Phys.\ Rev.\ Lett.\  {\bf 84}, 1398 (2000). 
 


\bibitem{Gayou:2001qd} 
  O.~Gayou {\it et al.} [Jefferson Lab Hall A Collaboration],
  Phys.\ Rev.\ Lett.\  {\bf 88}, 092301 (2002)
 


\bibitem{Punjabi:2005wq} 
  V.~Punjabi {\it et al.},
  Phys.\ Rev.\ C {\bf 71}, 055202 (2005)
  [Phys.\ Rev.\ C {\bf 71}, 069902 (2005)]. 
 


\bibitem{Puckett:2010ac} 
  A.~J.~R.~Puckett {\it et al.},
  Phys.\ Rev.\ Lett.\  {\bf 104}, 242301 (2010). 
 
\bibitem{Punjabi:2015bba} 
  V.~Punjabi, C.~F.~Perdrisat, M.~K.~Jones, E.~J.~Brash and C.~E.~Carlson,
  Eur.\ Phys.\ J.\ A {\bf 51}, 79 (2015). 
  
\bibitem{Guichon:2003qm} 
  P.~A.~M.~Guichon and M.~Vanderhaeghen,
  Phys.\ Rev.\ Lett.\  {\bf 91}, 142303 (2003). 


\bibitem{Blunden:2003sp} 
  P.~G.~Blunden, W.~Melnitchouk and J.~A.~Tjon,
  Phys.\ Rev.\ Lett.\  {\bf 91}, 142304 (2003). 


\bibitem{Carlson:2007sp} 
  C.~E.~Carlson and M.~Vanderhaeghen,
  Ann.\ Rev.\ Nucl.\ Part.\ Sci.\  {\bf 57}, 171 (2007). 
 


\bibitem{Arrington:2011dn} 
  J.~Arrington, P.~G.~Blunden and W.~Melnitchouk,
  Prog.\ Part.\ Nucl.\ Phys.\  {\bf 66}, 782 (2011).
 

\bibitem{DeRujula:1972te} 
  A.~De Rujula, J.~M.~Kaplan and E.~De Rafael,
  Nucl.\ Phys.\ B {\bf 35}, 365 (1971).

\bibitem{Zhang:2015kna} 
  Y.~W.~Zhang {\it et al.},
  Phys.\ Rev.\ Lett.\  {\bf 115}, no. 17, 172502 (2015).
  
\bibitem{Afanasev:2002gr} 
  A.~Afanasev, I.~Akushevich and N.~P.~Merenkov,
  hep-ph/0208260.
  
  
\bibitem{Gorchtein:2004ac} 
  M.~Gorchtein, P.~A.~M.~Guichon and M.~Vanderhaeghen,
  Nucl.\ Phys.\ A {\bf 741}, 234 (2004).

\bibitem{Pasquini:2004pv} 
  B.~Pasquini and M.~Vanderhaeghen,
  Phys.\ Rev.\ C {\bf 70}, 045206 (2004).
    
\bibitem{Kumar:2013yoa} 
  K.~S.~Kumar, S.~Mantry, W.~J.~Marciano and P.~A.~Souder,
  Ann.\ Rev.\ Nucl.\ Part.\ Sci.\  {\bf 63}, 237 (2013). 
  
\bibitem{Wells:2000rx} 
  S.~P.~Wells {\it et al.} [SAMPLE Collaboration],
  Phys.\ Rev.\ C {\bf 63}, 064001 (2001). 


\bibitem{Maas:2004pd} 
  F.~E.~Maas {\it et al.},
  Phys.\ Rev.\ Lett.\  {\bf 94}, 082001 (2005).

\bibitem{BalaguerRios:2012uk} 
  D.~Balaguer Rios,
  Nuovo Cim.\ C {\bf 035N04}, 198 (2012).
  
\bibitem{Armstrong:2007vm} 
  D.~S.~Armstrong {\it et al.} [G0 Collaboration],
  Phys.\ Rev.\ Lett.\  {\bf 99}, 092301 (2007). 
  
\bibitem{Androic:2011rh} 
  D.~Androic {\it et al.} [G0 Collaboration],
  Phys.\ Rev.\ Lett.\  {\bf 107}, 022501 (2011). 
  
  
\bibitem{Abrahamyan:2012cg} 
  S.~Abrahamyan {\it et al.} [HAPPEX and PREX Collaborations],
  Phys.\ Rev.\ Lett.\  {\bf 109}, 192501 (2012).

\bibitem{Waidyawansa:2013yva} 
  B.~P.~Waidyawansa [Qweak Collaboration],
  AIP Conf.\ Proc.\  {\bf 1560}, 583 (2013).
    
\bibitem{Nuruzzaman:2015vba} 
  Nuruzzaman [Qweak Collaboration],
  arXiv:1510.00449 [nucl-ex].



\bibitem{Rachek:2014fam} 
  I.~A.~Rachek {\it et al.},
  Phys.\ Rev.\ Lett.\  {\bf 114}, no. 6, 062005 (2015).


\bibitem{Adikaram:2014ykv} 
  D.~Adikaram {\it et al.} [CLAS Collaboration],
  Phys.\ Rev.\ Lett.\  {\bf 114}, 062003 (2015). 
  
  
\bibitem{Rimal:2016toz} 
  D.~Rimal {\it et al.} [CLAS Collaboration],
  arXiv:1603.00315 [nucl-ex].


\bibitem{Henderson:2016dea} 
  B.~S.~Henderson {\it et al.} [OLYMPUS Collaboration],
  Phys.\ Rev.\ Lett.\  {\bf 118}, no. 9, 092501 (2017).



\bibitem{Meziane:2010xc} 
  M.~Meziane {\it et al.} [GEp2gamma Collaboration],
  Phys.\ Rev.\ Lett.\  {\bf 106}, 132501 (2011). 

\bibitem{Guttmann:2010au} 
  J.~Guttmann, N.~Kivel, M.~Meziane and M.~Vanderhaeghen,
  Eur.\ Phys.\ J.\ A {\bf 47}, 77 (2011).




\bibitem{Pohl:2010zza} 
  R.~Pohl {\it et al.},
  Nature {\bf 466}, 213 (2010).


\bibitem{Antognini:1900ns} 
  A.~Antognini {\it et al.},
  Science {\bf 339}, 417 (2013).

\bibitem{Antognini:2013jkc} 
  A.~Antognini, F.~Kottmann, F.~Biraben, P.~Indelicato, F.~Nez and R.~Pohl,
  Annals Phys.\  {\bf 331}, 127 (2013). 


\bibitem{Carlson:2011zd} 
  C.~E.~Carlson and M.~Vanderhaeghen,
  Phys.\ Rev.\ A {\bf 84}, 020102 (2011).


\bibitem{Birse:2012eb} 
  M.~C.~Birse and J.~A.~McGovern,
  Eur.\ Phys.\ J.\ A {\bf 48}, 120 (2012). 
  

\bibitem{Pohl1:2016xoo} 
  R.~Pohl {\it et al.} [CREMA Collaboration],
  Science {\bf 353}, no. 6300, 669 (2016).
  
  

    
 \bibitem{Krauth:2015nja} 
  J.~J.~Krauth, M.~Diepold, B.~Franke, A.~Antognini, F.~Kottmann and R.~Pohl,
  Annals Phys.\  {\bf 366}, 168 (2016). 

\bibitem{Pineda:2004mx} 
  A.~Pineda,
  Phys.\ Rev.\ C {\bf 71}, 065205 (2005).

\bibitem{Nevado:2007dd} 
  D.~Nevado and A.~Pineda,
  Phys.\ Rev.\ C {\bf 77}, 035202 (2008).

\bibitem{Hagelstein:2015egb} 
  F.~Hagelstein, R.~Miskimen and V.~Pascalutsa,
  Prog.\ Part.\ Nucl.\ Phys.\  {\bf 88}, 29 (2016). 

\bibitem{Pineda:2002as} 
  A.~Pineda,
  Phys.\ Rev.\ C {\bf 67}, 025201 (2003).

\bibitem{Hill:2012rh} 
  R.~J.~Hill, G.~Lee, G.~Paz and M.~P.~Solon,
  Phys.\ Rev.\ D {\bf 87}, 053017 (2013).
  
\bibitem{Dye:2016uep} 
  S.~P.~Dye, M.~Gonderinger and G.~Paz,
  Phys.\ Rev.\ D {\bf 94}, no. 1, 013006 (2016).

\bibitem{Bernauer:2010wm} 
  J.~C.~Bernauer {\it et al.} [A1 Collaboration],
  Phys.\ Rev.\ Lett.\  {\bf 105}, 242001 (2010).
  
\bibitem{Bernauer:2013tpr} 
  J.~C.~Bernauer {\it et al.} [A1 Collaboration],
  Phys.\ Rev.\ C {\bf 90}, no. 1, 015206 (2014).

\bibitem{Mohr:2012tt} 
  P.~J.~Mohr, B.~N.~Taylor and D.~B.~Newell,
  Rev.\ Mod.\ Phys.\  {\bf 84}, 1527 (2012).
  
  
\bibitem{Bernauer:2014cwa} 
  J.~C.~Bernauer and R.~Pohl,
  Sci.\ Am.\  {\bf 310}, 32 (2014).
  
  \bibitem{Carlson:2015jba} 
  C.~E.~Carlson,
  Prog.\ Part.\ Nucl.\ Phys.\  {\bf 82}, 59 (2015). 
    
  
\bibitem{Mihovilovic:2016rkr} 
  M.~Mihovilovi\v{c} {\it et al.},
  arXiv:1612.06707 [nucl-ex].
  
\bibitem{Peng:2016szv} 
  C.~Peng and H.~Gao,
  EPJ Web Conf.\  {\bf 113}, 03007 (2016).
  
\bibitem{Gilman:2013eiv} 
  R.~Gilman {\it et al.} [MUSE Collaboration],
  arXiv:1303.2160 [nucl-ex].
  

\bibitem{Chen:2004tw} 
  Y.~C.~Chen, A.~Afanasev, S.~J.~Brodsky, C.~E.~Carlson and M.~Vanderhaeghen,
  Phys.\ Rev.\ Lett.\  {\bf 93}, 122301 (2004). 
  
\bibitem{Afanasev:2005mp} 
  A.~V.~Afanasev, S.~J.~Brodsky, C.~E.~Carlson, Y.~C.~Chen and M.~Vanderhaeghen,
  Phys.\ Rev.\ D {\bf 72}, 013008 (2005).


\bibitem{Borisyuk:2008db} 
  D.~Borisyuk and A.~Kobushkin,
  Phys.\ Rev.\ D {\bf 79}, 034001 (2009).
  
\bibitem{Kivel:2009eg} 
  N.~Kivel and M.~Vanderhaeghen,
  Phys.\ Rev.\ Lett.\  {\bf 103}, 092004 (2009). 
  
  
\bibitem{Kivel:2012vs} 
  N.~Kivel and M.~Vanderhaeghen,
  JHEP {\bf 1304}, 029 (2013). 


\bibitem{McKinley:1948zz} 
  W.~A.~McKinley and H.~Feshbach,
  Phys.\ Rev.\  {\bf 74}, 1759 (1948).



\bibitem{Brown:1970te} 
  R.~W.~Brown,
  Phys.\ Rev.\ D {\bf 1}, 1432 (1970).


\bibitem{Gorchtein:2014hla} 
  M.~Gorchtein,
  Phys.\ Rev.\ C {\bf 90}, no. 5, 052201 (2014).

\bibitem{Gorchtein:2006mq} 
  M.~Gorchtein,
  Phys.\ Lett.\ B {\bf 644}, 322 (2007). 

\bibitem{Tomalak:2015aoa} 
  O.~Tomalak and M.~Vanderhaeghen,
  Phys.\ Rev.\ D {\bf 93}, no. 1, 013023 (2016). 

\bibitem{Tomalak:2015hva} 
  O.~Tomalak and M.~Vanderhaeghen,
  Eur.\ Phys.\ J.\ C {\bf 76}, no. 3, 125 (2016).


\bibitem{Kondratyuk:2005kk} 
  S.~Kondratyuk, P.~G.~Blunden, W.~Melnitchouk and J.~A.~Tjon,
  Phys.\ Rev.\ Lett.\  {\bf 95}, 172503 (2005).

\bibitem{Borisyuk:2012he} 
  D.~Borisyuk and A.~Kobushkin,
  Phys.\ Rev.\ C {\bf 86}, 055204 (2012). 


\bibitem{Graczyk:2013pca} 
  K.~M.~Graczyk,
  Phys.\ Rev.\ C {\bf 88}, 065205 (2013).
  
  
\bibitem{Zhou:2014xka} 
  H.~Q.~Zhou and S.~N.~Yang,
  Eur.\ Phys.\ J.\ A {\bf 51}, no. 8, 105 (2015).

\bibitem{Lorenz:2014yda} 
  I.~T.~Lorenz, U.~G.~Mei\ss ner, H.-W.~Hammer and Y.-B.~Dong,
  Phys.\ Rev.\ D {\bf 91}, no. 1, 014023 (2015). 


\bibitem{Kondratyuk:2007hc} 
  S.~Kondratyuk and P.~G.~Blunden,
  Phys.\ Rev.\ C {\bf 75}, 038201 (2007).

\bibitem{Peset:2014jxa} 
  C.~Peset and A.~Pineda,
  Nucl.\ Phys.\ B {\bf 887}, 69 (2014).

\bibitem{Alarcon:2013cba} 
  J.~M.~Alarcon, V.~Lensky and V.~Pascalutsa,
  Eur.\ Phys.\ J.\ C {\bf 74}, no. 4, 2852 (2014). 


\bibitem{Borisyuk:2008es} 
  D.~Borisyuk and A.~Kobushkin,
  Phys.\ Rev.\ C {\bf 78}, 025208 (2008). 


  
\bibitem{Tomalak:2014sva} 
  O.~Tomalak and M.~Vanderhaeghen,
  Eur.\ Phys.\ J.\ A {\bf 51}, no. 2, 24 (2015).
  

\bibitem{Borisyuk:2013hja} 
  D.~Borisyuk and A.~Kobushkin,
  Phys.\ Rev.\ C {\bf 89}, no. 2, 025204 (2014).
 



\bibitem{Borisyuk:2015xma} 
  D.~Borisyuk and A.~Kobushkin,
  Phys.\ Rev.\ C {\bf 92}, no. 3, 035204 (2015).
  


\bibitem{Drechsel:1998hk} 
  D.~Drechsel, O.~Hanstein, S.~S.~Kamalov and L.~Tiator,
  Nucl.\ Phys.\ A {\bf 645}, 145 (1999).
  



\bibitem{Drechsel:2007if} 
  D.~Drechsel, S.~S.~Kamalov and L.~Tiator,
  Eur.\ Phys.\ J.\ A {\bf 34}, 69 (2007).
  
  
  
\bibitem{Pascalutsa:2005ts} 
  V.~Pascalutsa and M.~Vanderhaeghen,
  Phys.\ Rev.\ Lett.\  {\bf 95}, 232001 (2005). 
  
\bibitem{Pascalutsa:2005nd} 
  V.~Pascalutsa and M.~Vanderhaeghen,
  Phys.\ Lett.\ B {\bf 636}, 31 (2006). 
  
  
\bibitem{Pascalutsa:2006up} 
  V.~Pascalutsa, M.~Vanderhaeghen and S.~N.~Yang,
  Phys.\ Rept.\  {\bf 437}, 125 (2007). 


\bibitem{Jacob:1959at} 
  M.~Jacob and G.~C.~Wick,
  Annals Phys.\  {\bf 7}, 404 (1959)
  [Annals Phys.\  {\bf 281}, 774 (2000)].



\bibitem{Dennery:1961zz} 
  P.~Dennery,
  Phys.\ Rev.\  {\bf 124}, 2000 (1961).
 


\bibitem{Berends:1967vi} 
  F.~A.~Berends, A.~Donnachie and D.~L.~Weaver,
  Nucl.\ Phys.\ B {\bf 4}, 1 (1967).



\bibitem{Pasquini:2007fw} 
  B.~Pasquini, D.~Drechsel and L.~Tiator,
  Eur.\ Phys.\ J.\ A {\bf 34}, 387 (2007). 


\bibitem{Tomalak_PhD}
	O. Tomalak, dissertation, Johannes Gutenberg-Universit\"at Mainz, 2016.


\bibitem{Arrington:2007ux} 
  J.~Arrington, W.~Melnitchouk and J.~A.~Tjon,
  Phys.\ Rev.\ C {\bf 76}, 035205 (2007). 



\bibitem{Christy:2007ve} 
  M.~E.~Christy and P.~E.~Bosted,
  Phys.\ Rev.\ C {\bf 81}, 055213 (2010).



\bibitem{Donnachie:2004pi} 
  A.~Donnachie and P.~V.~Landshoff,
  Phys.\ Lett.\ B {\bf 595}, 393 (2004).

  
\end{thebibliography}
\end{document}